\documentclass[preprint2]{aastex}
\usepackage{graphics}

\shorttitle{Neutral Hydrogen in NGC~6503}
\shortauthors{Greisen, et al.}

\slugcomment{Draft of \today}

\begin{document}

\title{Aperture Synthesis Observations of the Nearby Spiral
    NGC~6503: Modeling the Thin and Thick HI Disks}

\author{Eric W. Greisen\altaffilmark{1}, Kristine
Spekkens\altaffilmark{2}\altaffilmark{3}, \and\ Gustaaf A. van
Moorsel\altaffilmark{1}}

\altaffiltext{1}{National Radio Astronomy Observatory, P.O. Box O,
  Socorro, NM 87801, U.S.A.}
\altaffiltext{2}{Department of Physics, Royal Military College of
  Canada, P.O. Box 17000, Stn Forces, Kingston ON, K7K 7B4}
\altaffiltext{3}{Previously a Jansky Fellow of the National Radio
  Astronomy Observatory.}

\begin{abstract}
We present sensitive aperture synthesis observations of the nearby,
late-type spiral galaxy NGC~6503, and produce HI maps of considerably
higher quality than previous observations by \citet{vMW85}.  We find
that the velocity field, while remarkably regular, contains clear
evidence for irregularities.  The HI is distributed over an area much
larger than the optical image of the galaxy, with spiral features in
the outer parts and localized holes within the HI distribution.  The
absence of absorption towards the nearby quasar 1748+700 yields an
upper limit of $5 \times 10^{17}\,\rm{cm}^{-2}$ for the column density
of cold HI gas along a line of sight which should intersect the disk at a
radius of 29 kpc. This suggests that the radial extent of the HI disk
is not much larger than that which we trace in HI emission (23 kpc).  The
observed HI distribution is inconsistent with models of a single thin
or thick disk.  Instead, the data require a model containing a thin
disk plus a thicker low column-density HI layer that rotates more
slowly than the thin disk and that extends only to approximately the
optical radius.  This suggests that the presence of extra-planar gas
in this galaxy is largely the result of 
star formation in the disk rather than cold gas accretion. Improved
techniques for interferometric imaging including multi-scale Clean
that were used in this work are also described.
\end{abstract}


\section{Introduction}
\label{s:intro}

Attention was drawn to the relatively nearby, low-luminosity spiral
galaxy NGC~6503 by the occurrence of a bright quasar only 5.3 arc
minutes away.  Deep optical plates taken by \citet{ASWdR76} show wisps
or spiral arms in the outermost portions of the NW part of the galaxy,
which were in the direction of, but did not approach, the quasar.
The neutral hydrogen was observed with the Westerbork Synthesis Radio
Telescope by \citet{SWC81} and later with the early, incomplete
VLA by \citet{vMW85}.  Neither group found any connection between the
quasar and the galaxy.  The images obtained by \citet{vMW85}
revealed a very regular rotation curve, but were sensitive only down
to a column density of $1.3 \times 10^{20}$ ${\rm cm}^{-2}$.

Although quasars are no longer thought to be directly linked to
nearby galaxies, NGC~6503 remains an interesting target: its
proximity, isolation, and regular kinematics make it ideal for
detailed structural studies to understand better the formation
and evolution of nearby spirals. We therefore undertook to re-observe
this galaxy using the full VLA with improved receivers in the C array
in order to obtain substantially better spatial and spectral
resolution with improved sensitivity.  Improvements in software
allowed us to take advantage of the bright quasar for calibration
without suffering the sidelobe problems encountered particularly by
\citet{SWC81} and further improvements allowed us to image the galaxy
with very little loss of fidelity due to missing short spacings.

Deep HI observations of nearby spirals have revealed a wealth of
low column density features at anomalous positions and velocities.
In particular, studies of systems such as NGC~2403 \citep{SSS00,
FMSO02}, NGC~2613 \citep{CI01}, NGC~4559 \citep{BFOBBS05}, NGC~891
\citep{OFS07}, UGC~7321 \citep{MW03}, NGC~253 \citep{BOFHS05},
NGC~6946 \citep{BOFHS08} and M83 \citep{MBW08} suggest that
extra-planar gas with decreasing rotation velocity with increasing
height from the disk may be ubiquitous in these systems (see
\citet{F08} for a complete list).  This anomalous gas is not clearly
associated with previous interactions or the presence of companions.
It has been proposed that this extra-planar HI stems from a galactic
fountain mechanism produced by feedback from supernovae in the disk
\citep[e.g][]{CBR02,FB06,OFS07, BOFHS08, MBW08}, although at least
some of it must stem from cold gas accretion from the intergalactic
medium \citep[e.g.][]{BCFS06,KMWSM06,FB08,SFOH08}.  It has proven
difficult to distinguish between these two possibilities, however, in
part because many of the galaxies imaged in HI are actively forming
stars.

NGC~6503 is an interesting system to examine in this context because
it offers the possibility of distinguishing between the star formation
and gas accretion hypotheses.  It has a relatively modest star
formation rate and an HI layer that extends far beyond the
star-forming disk.  If primordial gas accretion is a dominant
contributor to the extra-planar HI, one would expect little
correlation between this layer and the star-forming disk.  On the
other hand, if star formation were responsible, one wouldn't expect to
find a thick HI layer beyond the optical disk.

The data set presented in this paper provides sensitive HI maps of
NGC~6503\@.  Dynamical models of this galaxy obtained from these data
in combination with high-quality H$\alpha$ and CO kinematics are
forthcoming.  Here, we describe our HI observations and the derived
properties of NGC~6503 in preparation for that work, and model the
properties of its extra-planar HI.  In the appendix, we introduce
improved image deconvolution algorithms, describing in some detail the
particular implementation, widely available in the astronomical
community, used on the present data.

\subsection{Known properties of NGC~6503}
\label{se:known}

\begin{table*}
\centering
\renewcommand{\arraystretch}{1.1}
\caption{NGC~6503 properties \label{t:prop}}
\begin{tabular}{lll}
\noalign{\vspace{2pt}}
\tableline\tableline
parameter & value & reference \\
\tableline
Right ascension (J2000) & $17^{\rm h} 49^{\rm m} 26^{\rm s}.30$ 
                        & this work\\
Declination (J2000)     & $70^{\circ} 08' 40.7''$
                        & this work\\
Systemic Velocity (heliocentric) & $28.2 \pm 0.3$ km s$^{-1}$ & this work\\
Distance                & 5.2 Mpc     & \citet{KS97} \\
optical radius ($25^{\rm th}$ B mag/ss)
          & $3.55\arcmin$ & \citet{deV91} \\
optical radius          & 5.35 kpc & at 5.2 Mpc \\
Hubble type             & SA(s)cd     & \citet{deV91} \\
Absolute B magnitude    & -17.68 & \citet{M99} \\
B luminosity            & $1.5\times 10^9$ L$_\odot$ & \\
B central surface brightness & 20.05 mag arcsec$^{-2}$ & \citet{M99} \\
Star formation rate     & $0.18 \, \rm{M_{\odot}\,yr^{-1}}$ & this work\\
\tableline
\end{tabular}
\end{table*}

The basic properties of NGC~6503 are summarized in Table~\ref{t:prop}.
The galaxy is a late-type, low luminosity spiral located near the
Local Void \citep{Ketal03}.  Deep optical surveys have not found any
galaxies near NGC~6503, making it a very isolated, probably
unperturbed system. We note that because of its large angular size, it
is not included in standard isolated galaxy catalogs
\citep[e.g.][]{VM05}.

Due to its proximity, the distance to NGC~6503 cannot be determined
reliably from its recessional velocity.  Photometry of the brightest
stars in the galaxy led \citet{KS97} (see also \citet{Ketal03}) to
derive its distance as 5.2 Mpc, a value that we adopt in this study.
The Hubble type of SA(s)cd from \citet{deV91} indicates that NGC~6503
has an S shape and no bar, but \citet{LJLF07} prefer a Hubble type of
Sc rather than Scd.  \citet{M99} presents B, V, and I surface
brightness profiles traced to 200 arcsec radius, and finds the radius
of NGC~6503 at $25^{\rm th}$ magnitude per square arcsec to be 116.6,
140.6, and 198.9 arcsec, respectively at these bands.  The central
surface brightness of NGC~6503 lies in the range found for normal disk
galaxies.

We compute the global star formation rate of NGC~6503 using the
\citet{K98} IR star formation relation using the measurements of
\citet{SMKSS03}, following the method outlined by \citet{KGJD02}.
Its modest star formation rate explains its low B-band luminosity
compared to normal spirals of the same morphological type, even when
this quantity is normalized by the galaxy size and/or mass \citep{RH94}.

NGC~6503 is in general a fairly normal, if rather modest,
low-luminosity galaxy with very regular kinematics.  However, in the
center of this galaxy, \citet{LJLF07} find a bright central nucleus
which exhibits ``prominent LINER emission lines and a red continuum
with strong absorption features.''  Additionally, they find six point
sources of X-ray emission.  \citet{HFS97} call the nucleus a
``transition'' or even Seyfert type 2 with some uncertainty.  The
presence of some continuum radio emission in a larger, but central
region of NGC~6503 (Section~~\ref{images:cont}), suggests that
this galaxy hosts a low-luminosity active galactic nucleus (AGN).

\section{Observations and reductions}
\label{S:obs}

NGC~6503 was observed on 27 February 1996 with the VLA in C
configuration.  There were 127 spectral channels separated by 24.4 kHz
(5.15 km s$^{-1}$) centered on 26 km s$^{-1}$ heliocentric radial
velocity.  The sources used for bandpass shape and absolute flux
density calibration were 3C286 and 3C48, and were observed at the
beginning, middle, and end of the session for a total of about 43
minutes. The total integration time on NGC~6503 was a little more than
500 minutes, interspersed every 50 minutes or less with 2-minute
observations of the 2.0-Jy amplitude and phase calibrator 1800+784\@.
The observing setup is summarized in Table~\ref{t:obs}.

All data reduction was done in the AIPS software package
\citep{Gre03}.  After deleting $\sim2\%$ of the data for various
reasons such as scan start-up problems and spurious amplitudes and
phases, the amplitude and phase calibrations were applied to NGC~6503
(and 1800+784)\@.  The bandpass shapes were determined from the 3C286
and 3C48 observations, interpolated in time, and applied to NGC~6503
and 1800+784\@. Five channels exhibiting weak Galactic HI absorption
were flagged in 3C48, and the bandpass was interpolated across them
before being applied to the data.

The 0.76-Jy quasar 1748+700, in the same beam as NGC~6503, dominates
the data in all continuum channels.  To separate the continuum and
line signals, the data for NGC~6503 were phase shifted to the quasar
position and straight lines (with slope) were fit to the spectrum of
the real and imaginary parts of each visibility sample using a total
of 50 channels from the band edges which were judged to be both free
of line signal and bandpass edge effects.  A line-only dataset was
produced  by subtracting the fits from the visibility data and
shifting back to the original phase center, while the best-fitting
continuum values at the band center were adopted as the continuum-only
dataset.  A task similar to the one used is described by
\citet{CUH92}. 

All imaging and deconvolution were done with the AIPS program IMAGR
using a tempered uniform weighting (``robust weighting'',
\citet{Briggs99}).  The point-source response was nearly circular.  We
used IMAGR's multi-scale Clean algorithm to deconvolve the synthesized
beam from both the continuum and line datasets.  Because this
functionality has not yet been thoroughly described in the literature,
we discuss it in detail in Appendix~\ref{ap:imagr}.  The final image
properties are summarized in Table~\ref{t:obs}.

The continuum was imaged over an area approximately $2.5\arcdeg$ in
diameter in order to cover both the primary beam and the first outer
sidelobe of the individual VLA antennas.  A few more distant regions
surrounding nearby NVSS sources \citep{NVSS98} were also included.
The area was imaged in 55 facets \citep{CP92} in order to avoid
wide-field distortions.  The deconvolution was performed with a
multi-scale Clean using Gaussian model sources with full-width
half-maximum (FWHM) sizes of 0\arcsec, 36\arcsec, and 108\arcsec, and
Cleaning was stopped when the peak residual in each facet fell below a
flux density cutoff of 0.21, 0.42, and 1.7 mJy/beam, respectively.
The ``bias" parameter controlling the facet Cleaning order was
$b=0.62$ (see Appendix~\ref{ap:imagr}).

The proximity to NGC~6503 of the bright quasar 1748+700, as well as
other continuum sources, allows for an accurate self-calibration to
improve the amplitude and phase calibration, as initially described
by \citet{S80}.  The implementation in AIPS uses the full multi-scale,
multi-facet Clean model to compute model visibilities which are
divided into observed visibilities to get apparent baseline-dependent
gains.  A least-squares method turns these into antenna-dependent
gains which are applied to the observations before another round of
imaging.  Given the quality of the initial calibration using
1800+784 and the strength of 1748+700, the self-calibration
iterations converged quite rapidly.  After imaging, deconvolution
and several rounds of self-calibration, the 55 separate facet images
were interpolated onto a ``flattened'' image with no loss of intensity
or positional accuracy.  The final continuum image has a resolution of
14\arcsec, a root mean square (RMS) of $80\,\mu$Jy/beam and a dynamic
range in excess of 9000:1\@.

The improved amplitude and phase solutions from the self-calibration
and some additional editing were then applied to the HI visibility
data.  Each channel was imaged in a single $50\arcmin \times
50\arcmin$ facet. The deconvolution was performed with a multi-scale
Clean using Gaussian model sources with FWHMs of 0\arcsec, 36\arcsec,
108\arcsec\ and 324\arcsec, flux density cutoffs of 0.2, 0.55, 2.6,
and 7.8 mJy/beam respectively, and a bias parameter $b=0.62$ (see
Appendix~\ref{ap:imagr}).  The Clean was limited interactively to the
regions which clearly had emission in that channel in order to
mitigate the well-known Clean bias \citep{NVSS98}. The final HI data
cube has an RMS noise of 0.58 mJy/beam, a spatial resolution of
14\arcsec\ per channel, and a peak dynamic range of 30:1.

\begin{figure}
\centering
\resizebox{!}{2.9in}{\includegraphics{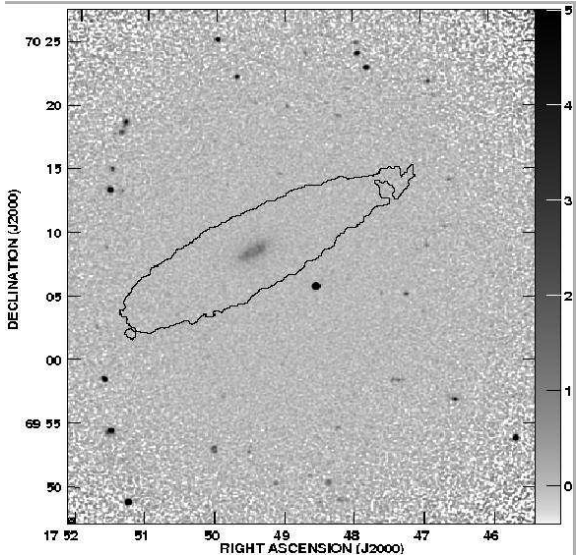}}
\caption{Primary beam-corrected continuum image of a field near 
  NGC~6503\@.  The wedge to the right shows the displayed brightness
  in mJy/beam which is clipped at 5 mJy/beam.  The extent of the
  detected HI emission is shown by the solid line (see
  Section~\ref{analysis:totHI}).  The quasar 1748+700 is the strong
  point source $5.34\arcmin$ to the SW of the galaxy.}
\label{fig:PBcor1}
\end{figure}

\begin{figure}
\centering
\resizebox{2.9in}{!}{\includegraphics{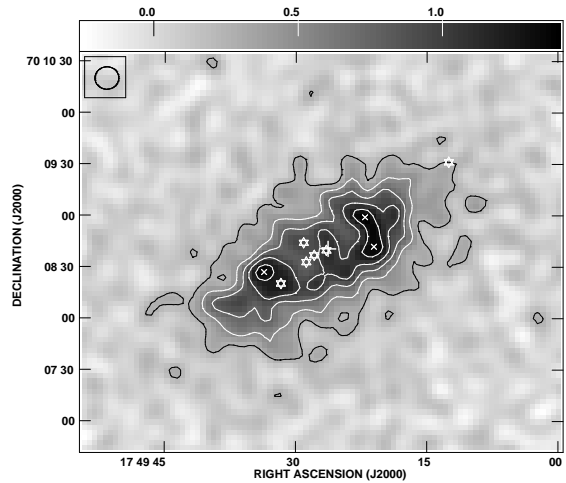}}
\caption{21-cm radio continuum distribution in NGC~6503\@.  The
  wedge at the top shows the displayed brightness in mJy/beam, and the
  contours are drawn at (1, 2, 3, 4, 5) $\times$ 0.25 mJy/beam. The
  Clean beam is plotted in the top left corner.  The continuum peaks
  are marked by crosses, the 6 unresolved X-ray sources found by
  \citet{LJLF07} are marked by stars, and the kinematic center of the
  HI distribution (Section~\ref{analysis:kin}) is marked by a plus
  sign.}
\label{fig:Ccent}
\end{figure}

\section{Images}
\label{images}

\begin{table}
\centering
\renewcommand{\arraystretch}{1.1}
\caption{Observing setup and image properties\label{t:obs}}
\begin{tabular}{lll}
\noalign{\vspace{2pt}}
\tableline\tableline
parameter & value & units \\
\tableline
VLA array configuration     & C    & \\
Primary beam FWHM           & 29.7 & arc minutes \\
Time on-source              & 500  & min. \\    
Usable total bandwidth      & 2513 & kHz \\
Band center (heliocentric)  & 26   & km s$^{-1}$ \\
Spatial resolution          & 14   & arcsec  \\
Spatial resolution          & 0.35 & kpc \\
Effective cont.~bandwidth   & 1220 & kHz \\
RMS image noise (cont.)     & 0.08 & mJy/beam \\
Dynamic range (cont.)       & 9000 & \\ 
Channel separation (line)   & 5.15 & km s$^{-1}$ \\
RMS image noise (line)      & 0.58 & mJy/beam \\
Peak dynamic range (line)   & 30 & \\ 
\tableline
\end{tabular}
\end{table}

The properties of the full-resolution continuum and HI images of
NGC~6503 are summarized in Table~\ref{t:obs}.  Below, we discuss
each in turn.

\subsection{Continuum in NGC~6503 and in 1748+700}
\label{images:cont}

The total Cleaned flux density from the $2.5\arcdeg$ diameter
continuum image in the vicinity of NGC~6503 is 967 mJy, uncorrected 
for attenuation from the primary beam of the individual VLA antennas.
There are $\sim 90$ sources with a signal-to-noise ratio $S/N > 5$ in
the image, some of which are in the first and second outer sidelobes
of the primary beam.  The sources are essentially all unresolved
except for NGC~6503\@.

Fig.~\ref{fig:PBcor1} presents a portion of the primary beam-corrected
image of the continuum emission in the vicinity of NGC~6503\@.  As
expected, the noise in the image rises rapidly towards the edge of the
primary beam.  The total image size is governed by the beam correction
itself, since the VLA's primary beam is completely unknown beyond
30\arcmin\ at this frequency.  The paucity of continuum sources near
NGC~6503 and 1748+700 as well as the apparent chain of sources east of
the pair is initially striking, but not statistically significant (J.
J. Condon, priv.~comm.).  The extent of the HI emission in NGC~6503
(see Section~\ref{analysis:totHI}) is represented by the solid line in
Fig.~\ref{fig:PBcor1}.  Qualitatively, the centroids and position
angle of the HI and continuum emission are in good agreement, with the
HI extending well beyond the continuum.

The continuum properties of NGC~6503 are summarized in
Table~\ref{t:results}.  The primary-beam corrected total flux density
of the extended continuum source in NGC~6503 is $39 \pm 1.2$ mJy.  A
contour image of the continuum radio source in NGC~6503 is shown in
Fig.~\ref{fig:Ccent}.  The position angle is close to the -60 degrees
found in the HI\@.  The full extent along the major axis is about
200\arcsec or 5 kpc at the adopted distance.  The continuum appears to
peak at the three locations marked by crosses.  The HI is at a minimum
in the center, but has peaks near the other three locations.  The
apparent ridge in the continuum in the NW crosses a region in which
the HI is minimal, suggesting that this correspondence is probably not
causal.  The six unresolved X-ray sources found by \citet{LJLF07} are
also illustrated with stars.  That the central continuum radio peak is
coincident with an X-ray source is further evidence for a
low-luminosity AGN \citep{HFS97}.

With primary beam correction, the total flux density of the quasar
1748+700 was 755 mJy on the observation date.  Within reasonable
uncertainties the quasar is unresolved.

\subsection{HI in NGC~6503}
\label{images:HI}

Fig.~\ref{fig:Line4} shows the detected HI emission from NGC~6503 in a
selection of channels.  It is clear that even at the higher resolution
and sensitivity of our data relative to those of \citet{vMW85}, the HI
kinematics of NGC~6503 are remarkably regular.  Nonetheless, when the
channel images are examined in detail, irregularities are apparent.
The outer portions of many emission ``wings'' bend away from the major
axis, most noticeably at -25.5 km s$^{-1}$, -10.1 km s$^{-1}$ and 82.7
km s$^{-1}$ in Fig.~\ref{fig:Line4}.  There are also spatially
distinct features in the distribution of some channels, such as at
-25.5 km s$^{-1}$ and 98.2 km s$^{-1}$.  We investigate the HI
morphology and kinematics of NGC~6503 in detail in
Section~\ref{analysis}.

\begin{figure*}
\centering
\resizebox{!}{7.25in}{\includegraphics{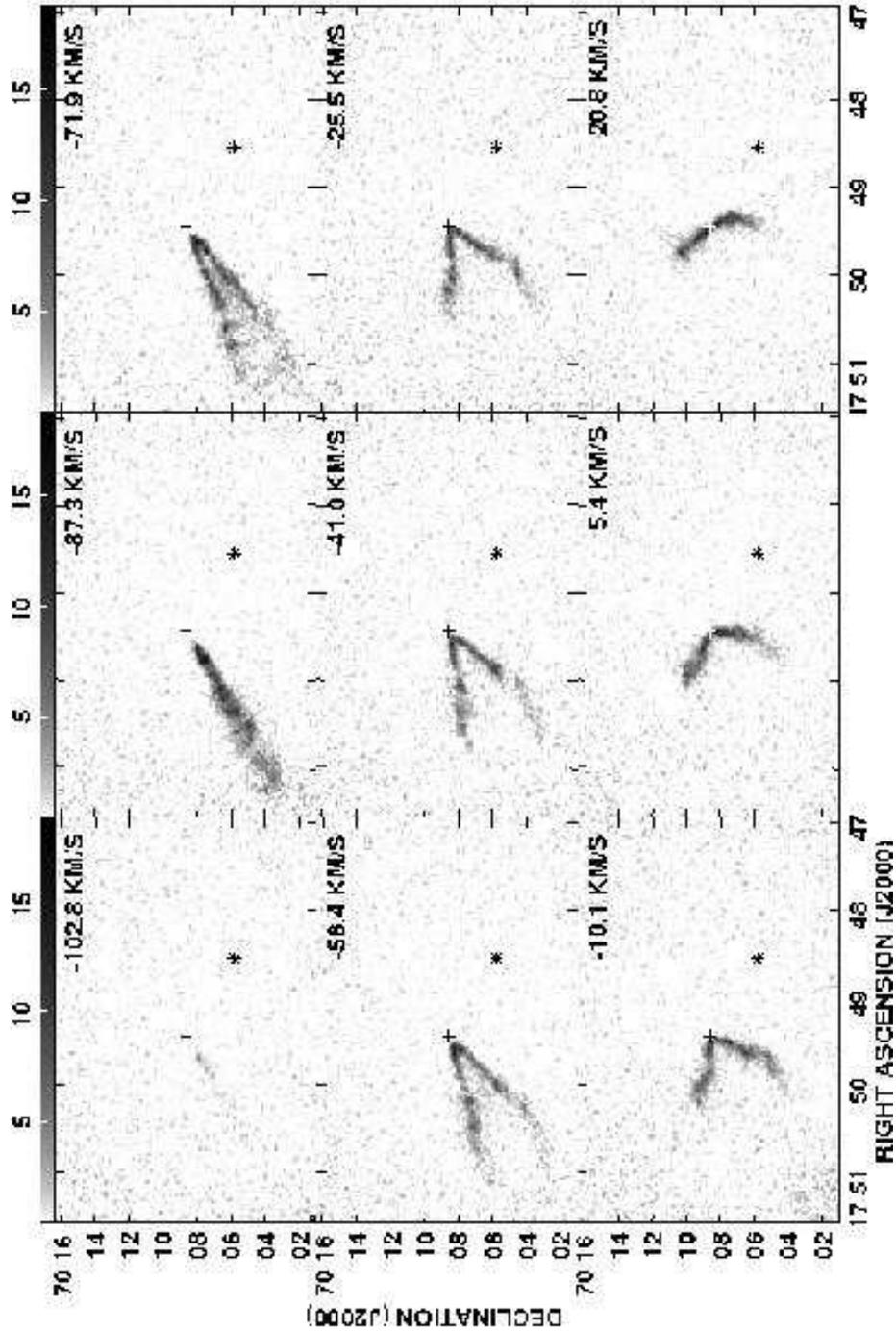}}
\caption{Primary-beam corrected channel images of the HI emission
  in NGC~6503\@.  The gray-scale is plotted from 0.5 to 18.65 mJy/beam
  with a logarithmic transfer function and contours are plotted at
  -2, 2, 4, 8, and 16 mJy/beam.  (The negative contours appear only in
  regions which are noise dominated.)  The heliocentric radial
  velocity of the channel is in the upper right corner, and the Clean
  beam is plotted in the lower left corner of the first panel.  The
  kinematic center of NGC~6503 is marked with a plus sign and the
  quasar 1748+700 is marked with an asterisk.}
\label{fig:Line4}
\end{figure*}

\begin{figure*}
\centering
\figurenum{3}
\resizebox{!}{7.4in}{\includegraphics{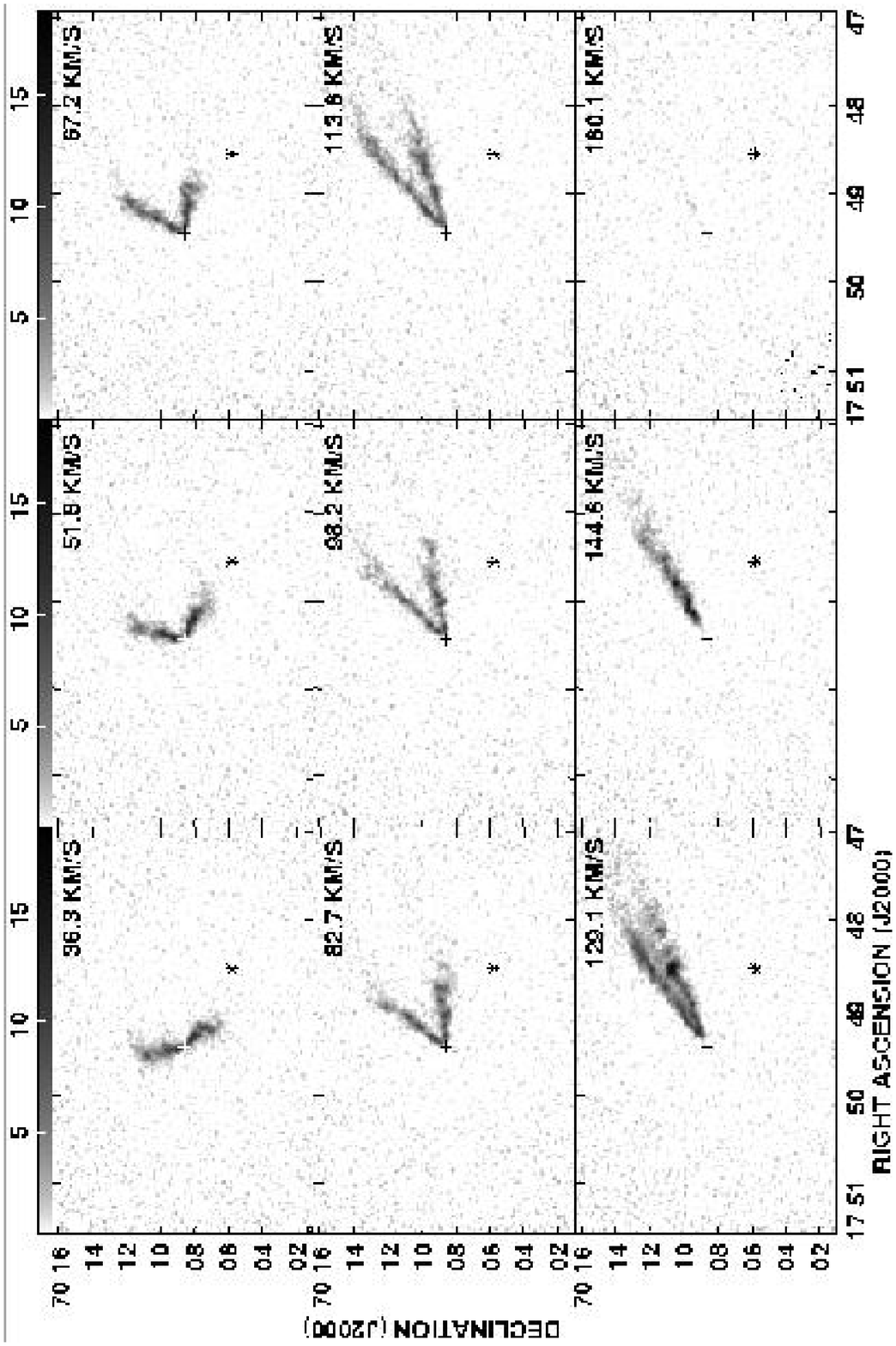}}
\caption{\it Continued.}
\label{fig:Line5}
\end{figure*}

\begin{figure*}
\centering
\resizebox{!}{7.4in}{\includegraphics{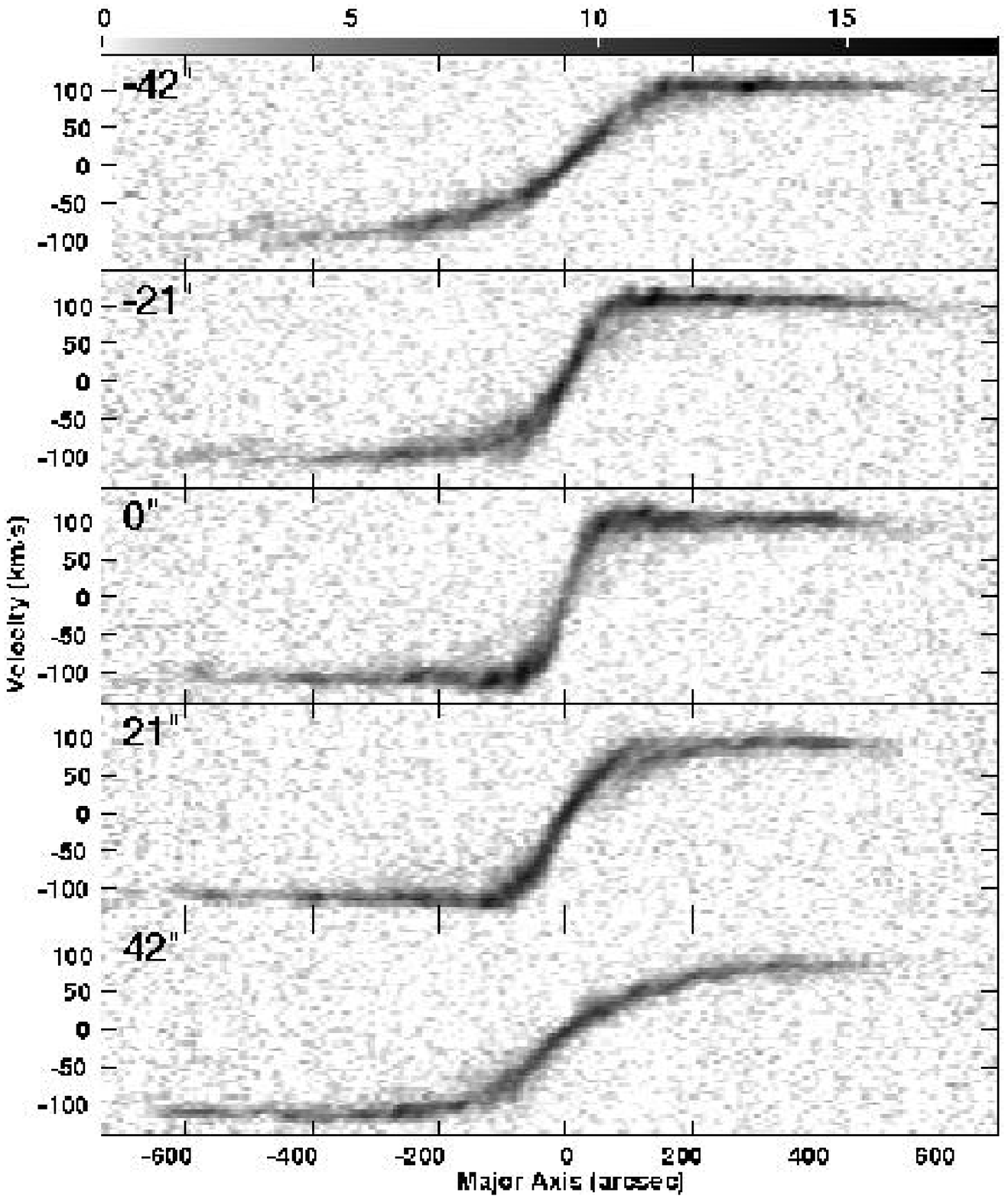}}
\caption{Primary-beam corrected images for slices parallel to the
  major axis, separated by 21\arcsec.  Axis labels are relative to the
  kinematic center and the systemic velocity.  The gray-scale is
  plotted from 0.5 to 18.65 mJy/beam with a logarithmic transfer
  function and contours are plotted at 1.5, 3, 6, and 12 mJy/beam.}
\label{fig:MajAx}
\end{figure*}

\begin{figure*}
\centering
\resizebox{!}{7.5in}{\includegraphics{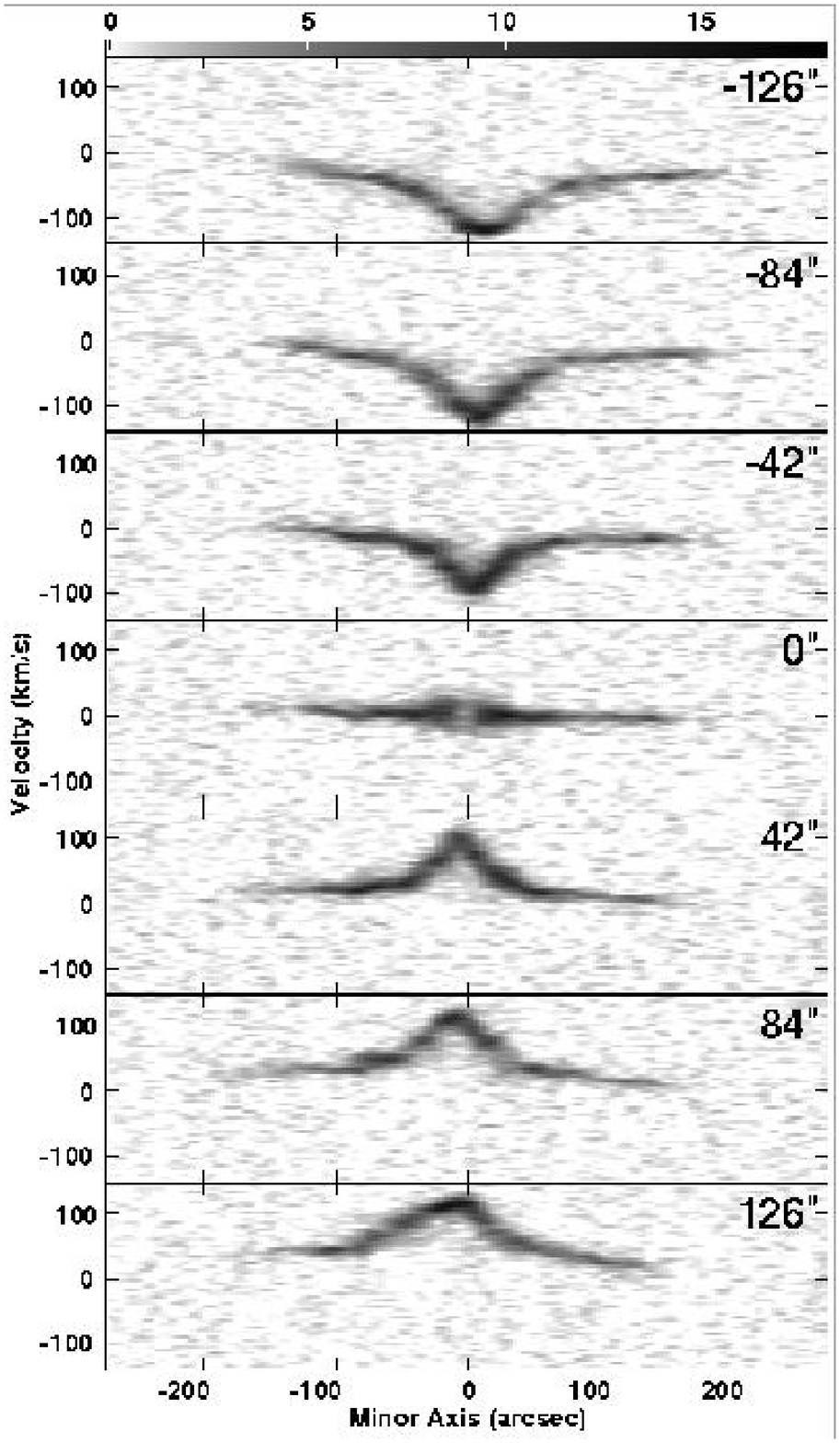}}
\caption{Primary-beam corrected images for slices parallel to the
  minor axis, separated by 42\arcsec.  Axis labels are relative
  to the kinematic center and the systemic velocity.  The gray-scale
  is plotted from 0.5 to 18.65 mJy/beam with a logarithmic transfer
  function and contours are plotted at 1.5, 3, 6, and 12 mJy/beam.}
\label{fig:MinAx}
\end{figure*}

Fig.~\ref{fig:MajAx} shows selections from the data cube after
rotation and transposition, plotting the HI as a function of velocity 
and position along lines parallel to the major axis.  The central
panel is the traditional rotation curve image of this galaxy.
Fig.~\ref{fig:MinAx} shows position-velocity images along the minor
axis at several positions on the major axis.  These figures are made
from the full resolution image using a single 3-arcsecond pixel on the
third axis for each panel with no additional smoothing.  At first
glance, the position-velocity slices of NGC~6503 exhibit considerable
symmetry and do not appear to show an extensive ``beard'' of HI
emission at velocities between the rotation velocity and the systemic
velocity, as found for several other systems
\Citep{FMSO02,MW03,BOFHS05,OFS07}.  In Section~\ref{beard}, we place
quantitative constraints on the morphology and kinematics of the
extra-planar HI layer that is consistent with the present data.  The
bottom panel in Fig.~\ref{fig:MinAx} in particular shows some
significant asymmetry.  This is associated with the warp to the SW
discussed further in Section~\ref{analysis:CUBIT} in connection with
Fig.~\ref{fig:CubitDiff}.

\subsection{HI in the vicinity of NGC~6503}

We searched the data cube outside the confines of NGC~6503 for any
other source of line emission.  Due to limitations of the VLA
correlator and the dishes that make up the array, a bandwidth of only
about 500 km s$^{-1}$ was observed,  centered at 26 km s$^{-1}$
heliocentric, over a region about 50\arcmin in diameter.  No
other extragalactic HI emission was found. For an unresolved source
spanning two contiguous velocity channels at the distance of NGC~6503,
this non-detection corresponds to a $3\sigma$ HI mass limit that
varies between $M_{HI} = 1.1 \times 10^5 \,M_\odot$ at the pointing
center and ten times this value near the edge of the field.  In the
limited volume probed by our observations, NGC~6503 appears isolated
from an HI standpoint as well as from an optical one.

\begin{figure}
\centering
\resizebox{3.0in}{!}{\includegraphics{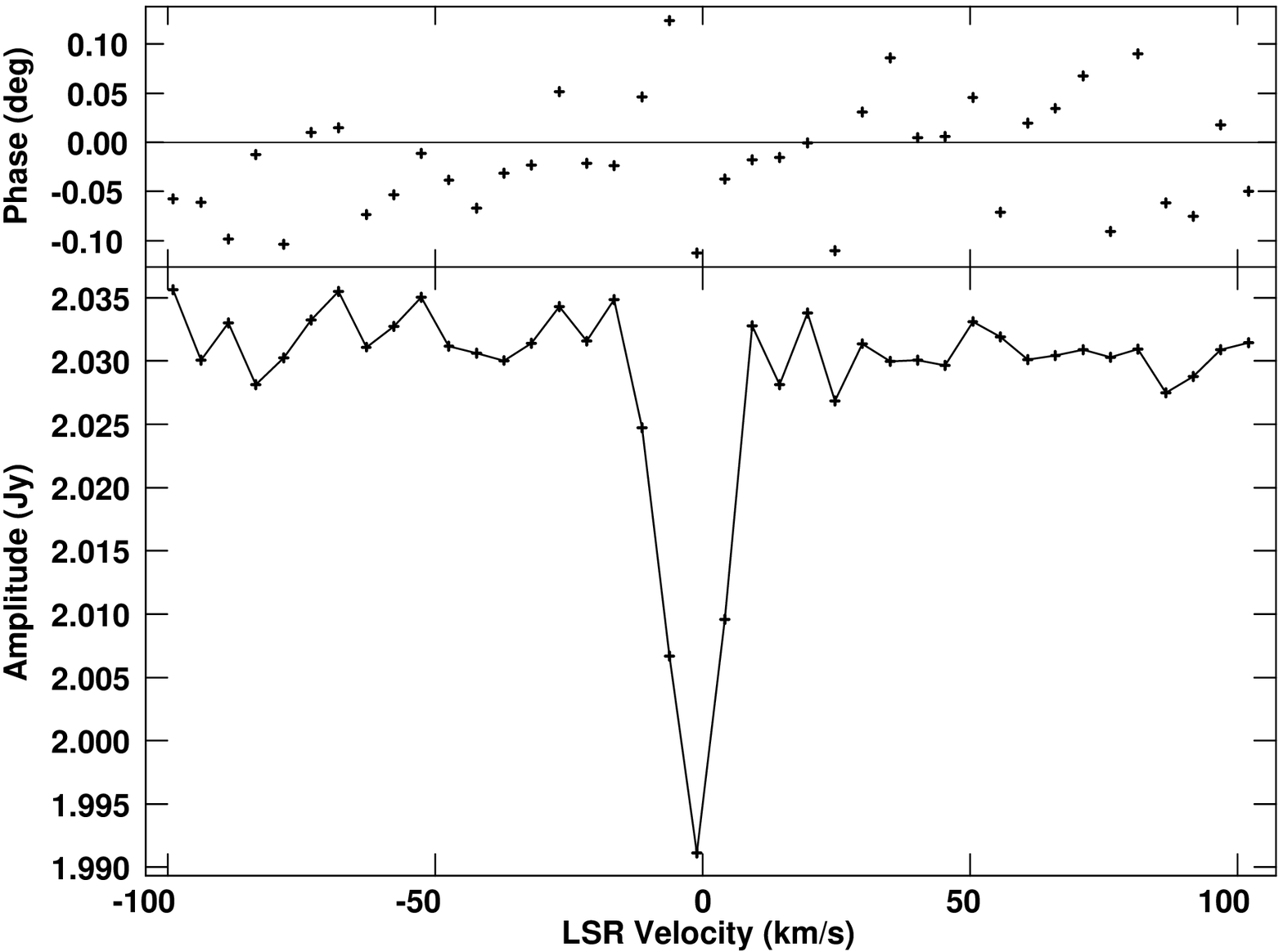}}
\caption{The spectrum of calibration source 1800+784 computed by
  vector averaging all calibrated fringe visibilities.}
\label{fig:1800pos}
\end{figure}

\begin{figure}
\centering
\resizebox{3.0in}{!}{\includegraphics{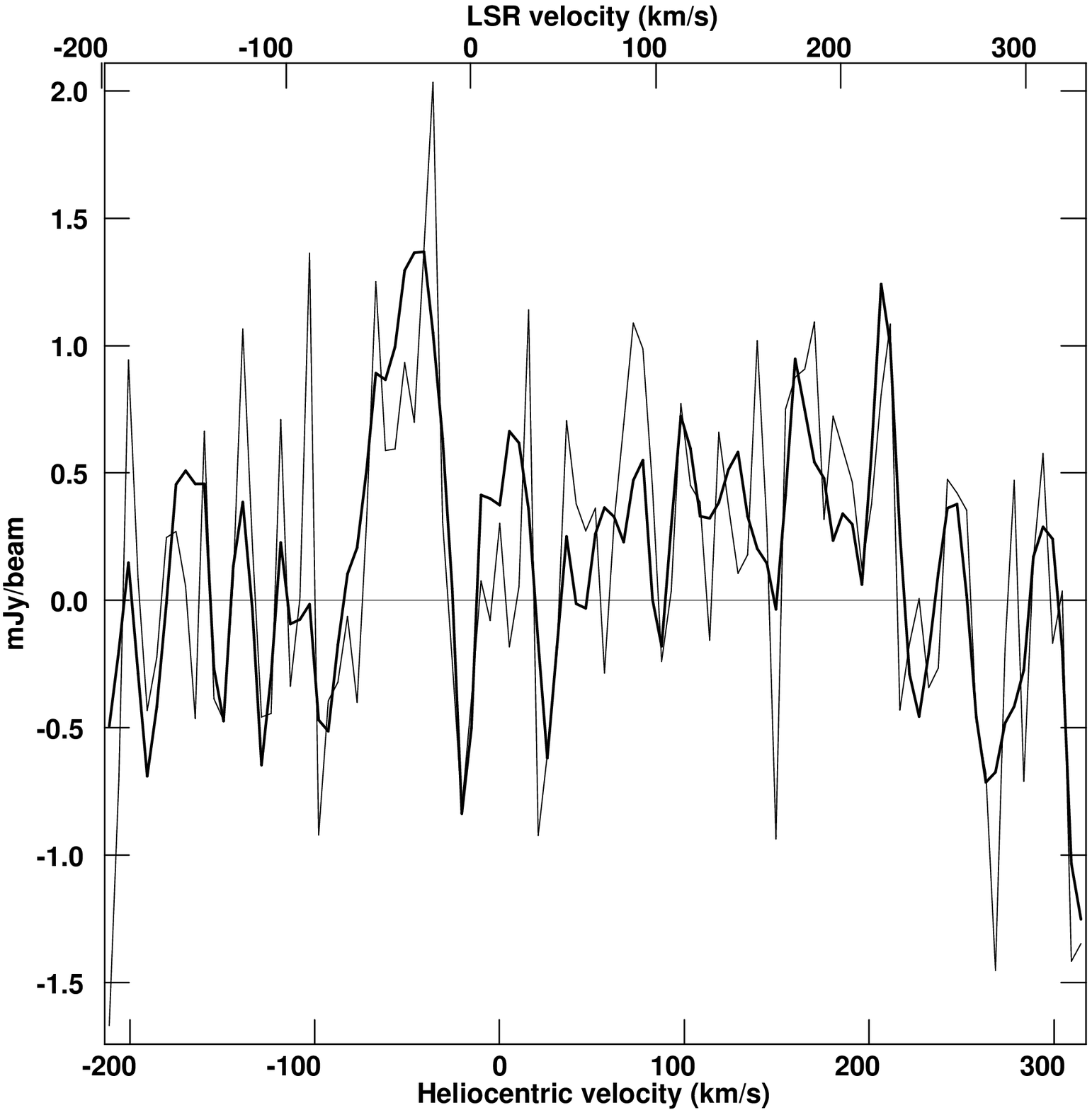}}
\caption{The spectrum at the position of 1748+700 (RA = 
  $17^\mathrm{h}48^\mathrm{m}32\fs84$, Dec =
  $+70\arcdeg05\arcmin50\farcs8$ in J2000) minus the measured
  continuum flux density of 755 mJy.  Light line from the
  high-resolution, primary-beam corrected data cube (RMS = 0.58
  mJy/beam).  Heavy line from a version of the data cube smoothed to
  30\arcsec\ spatial resolution and 10 km s$^{-1}$ velocity
  resolution (RMS = 0.38 mJy/beam).}
\label{fig:QuaSpec}
\end{figure}

\subsection{Absorption towards 1800+784 and 1748+700}

The calibration source 1800+784 is located at Galactic longitude
$l=110.0\arcdeg$, latitude $b=29.1\arcdeg$, while NGC~6503 is at
$l=100.5\arcdeg$, $b=30.7\arcdeg$.  These directions are expected to
have local Milky Way HI emission and, most likely, absorption.
Indeed, Fig.~\ref{fig:1800pos} shows an absorption line in the
spectrum of 1800+784, with a depth of 0.040 $\pm$ 0.002 Jy or an
optical depth of 0.02.  The line is centered at an LSR velocity of
-1.6 km s$^{-1}$ with a FWHM of 13 km s$^{-1}$.

By contrast, Fig.~\ref{fig:QuaSpec} shows that no absorption is
observed in the direction of the quasar 1748+700 ($l=100.5\arcdeg$,
$b=30.8\arcdeg$) to very high precision.  The deepest negatives
in the full resolution image in this direction (light line) are less
than 1 mJy/beam, while the spectrum extracted from an image smoothed
to 30\arcsec\ and 10 km s$^{-1}$ resolution (heavy line) shows even
less evidence for absorption.  The optical depth toward 1748+700 is
thus, conservatively, $\tau = 0.0 \pm 0.001$.  This suggests that, for
HI of line width 5 km s$^{-1}$ and temperature 50 K, the column
density upper limit is $5 \times 10^{17} {\rm cm}^{-2}$.  Assuming a
thin, flat disk in NGC~6503, the line of sight to the quasar passes
through the plane of NGC~6503 at a radius of 29 kpc.  There must be
very little cold HI at this radius in NGC~6503.  The absence of
absorption from the Milky Way is also remarkable.

We note that in Fig.~\ref{fig:QuaSpec}, the peak at an LSR velocity of
$\sim -30$ km s$^{-1}$ with a FWHM $\sim$ 40 km s$^{-1}$ coincides
with the channels contaminated by Milky Way emission over much of the
field.  The projected heliocentric velocity of the HI disk in NGC~6503
along this line of sight is $\sim 50$ km s$^{-1}$ (LSR velocity of
$\sim 65$ km s$^{-1}$); we therefore do not expect Milky Way emission
or absorption to confound an absorption signature from NGC~6503.

\section{HI in NGC~6503: analysis}
\label{analysis}

\begin{table*}
\centering
\renewcommand{\arraystretch}{1.1}
\caption{Measured properties of NGC~6503 \label{t:results}}
\begin{tabular}{lrlr}
\noalign{\vspace{2pt}}
\tableline\tableline
parameter & value & units & Section\\
\tableline
Continuum flux density             & $39.4 \pm 0.7$ & mJy & \ref{images:cont} \\
Continuum angular diameter         & $200 \pm 20$ & arcsec & \ref{images:cont} \\
Continuum linear diameter          & $5 \pm 0.5$ & kpc & \ref{images:cont} \\
Line integral                      & $205\pm 1$ & Jy km s$^{-1}$ & \ref{analysis:totHI} \\
HI mass                            & $1.3\pm 0.2$ & $ 10^9\, {\rm M}_\odot$ & \ref{analysis:totHI} \\
HI mass / B luminosity             & $0.87 \pm 0.13$ & ${\rm M}_\odot / {\rm L}_\odot$ & \ref{analysis:totHI}, \ref{se:known} \\
Integrated line width at 20\%\ peak & 246 & km s$^{-1}$ & \ref{analysis:totHI} \\
Integrated line width at 50\%\ peak & 234 & km s$^{-1}$ & \ref{analysis:totHI} \\
HI radius ($0.1\, {\rm M}_\odot \, /\, {\rm pc}^2$) NW & 22.6 & kpc & \ref{analysis:totHI} \\
HI radius ($0.1\, {\rm M}_\odot \, /\, {\rm pc}^2$) SE & 17.2 & kpc & \ref{analysis:totHI} \\
Mean HI/optical radius ratio       & 3.7  &  \\
Kinematic center (RA, J2000)       & $17^{\rm h} 49^{\rm m} 26^{\rm s}.30$ & & \ref{analysis:kin} \\
Kinematic center (Dec, J2000)      & $70^{\circ} 08' 40.7''$ & & \ref{analysis:kin} \\
Kinematic Position angle           & -60.1 & degrees & \ref{analysis:kin} \\
Inclination                        & 75.1  & degrees & \ref{analysis:CUBIT} \\
Systemic velocity (heliocentric) & $28.5 \pm 0.3$ & km s$^{-1}$ & \ref{analysis:kin} \\
dynamical mass ($R = R_{\rm max}$= 3.5 kpc) & 1.0  & $10^{10}\, {\rm M}_\odot$ & \ref{analysis:kin} \\
dynamical mass ($R = 6\,$kpc)      & 1.8  & $10^{10}\, {\rm M}_\odot$ & \ref{analysis:kin} \\
dynamical mass ($R = 20\,$kpc)     & 6.0  & $10^{10}\, {\rm M}_\odot$ & \ref{analysis:kin} \\
Max.~column density\tablenotemark{a} & 6.9  & ${\rm M}_\odot \, /\, {\rm pc}^2$ & \ref{analysis:CUBIT} \\
Max.~column density\tablenotemark{a} & 8.6 & $10^{20}$ atoms cm$^{-2}$ & \ref{analysis:CUBIT} \\
Minimum detectable signal         & 3 & Jy/beam m s$^{-1}$ & \ref{analysis:totHI} \\
Minimum detectable column density & $1.8 \times 10^{19}$ & cm$^{-2}$ & \ref{analysis:totHI} \\
\tableline
\end{tabular}
\tablenotetext{a}{corrected for inclination ($75.1^{\circ}$)}
\end{table*}

In this section we examine the HI morphology and kinematics of
NGC~6503 in detail.

Analysis of an image over an area much larger than that containing
real signal produces large uncertainties due to the inclusion of
regions containing only  noise.  To mitigate this effect in NGC~6503,
we produced a cube in which line-free regions were blanked.  The
multi-scale Cleaned image cube was smoothed in frequency with a
Gaussian of 3 channels FWHM and convolved to a spatial resolution of
25\arcsec, producing an image with RMS 0.22 mJy/beam.  Then all pixels
in the full resolution cube for which the smoothed cube was less in
absolute value than three times the RMS of the smoothed cube were
blanked. An interactive display program was used to eliminate the
remaining pixels throughout the cube that were disconnected from each
other and from any region of obvious line signal.

\subsection{Total HI and its distribution}
\label{analysis:totHI}

\begin{figure*}
\centering
\resizebox{5.in}{!}{\includegraphics{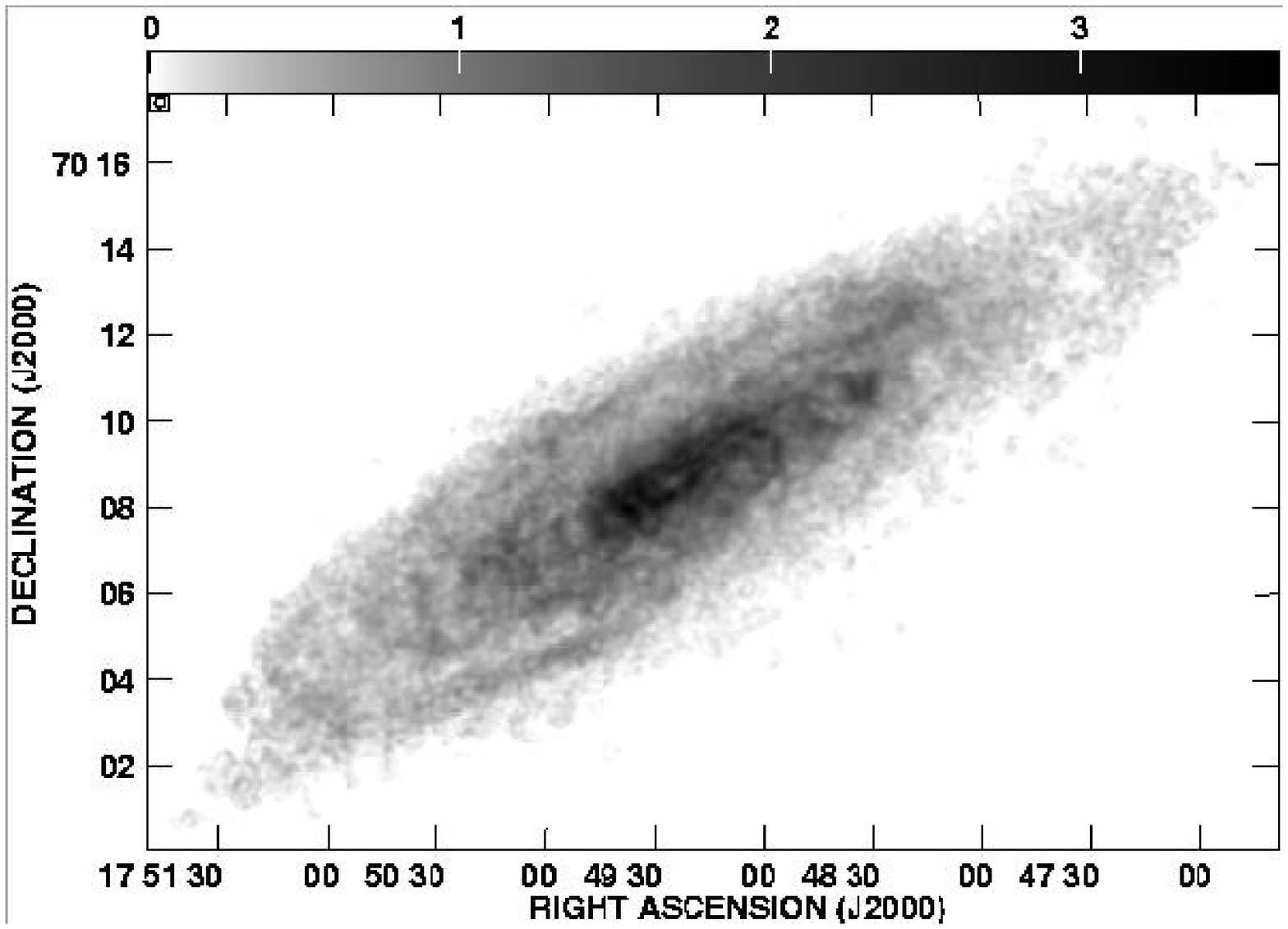}}
\vspace{12pt}
\resizebox{5.in}{!}{\includegraphics{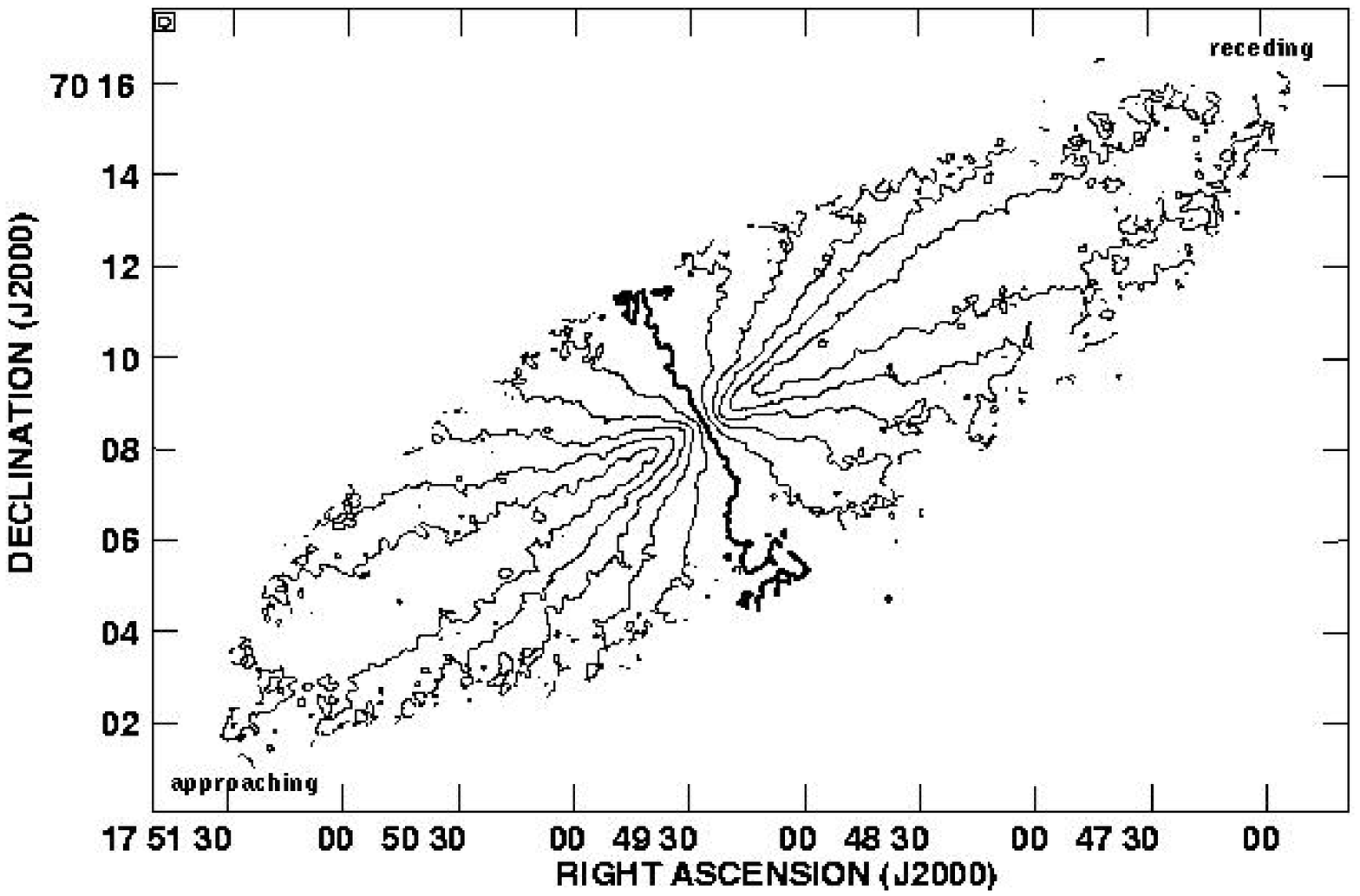}}
\caption{Top: zeroth moment of the HI distribution in NGC~6503
  measured in $10^{21}\, {\rm cm}^{-2}$ and plotted with a logarithmic
  gray-scale.  The column density is $0.57\times 10^{19}\,{\rm
    cm}^{-2}\,SdV$, with $SdV$ in units of Jy/beam m s$^{-1}$.)
  The lowest reliable HI emission from NGC~6503 is at $1.8 \times
  10^{19}\,{\rm cm}^{-2}$.  Bottom: first moment of the HI
  distribution in NGC~6503.  The heavy contour is plotted at 26 km
  s$^{-1}$ heliocentric (nearly the systemic velocity), and the other
  contours are spaced at 20 km s$^{-1}$ relative to this value.  The
  eastern (E) side is approaching, while the western (W) side is
  receding.}
\label{fig:Xmoms}
\end{figure*}

\begin{figure}
\centering
\resizebox{3.0in}{!}{\includegraphics{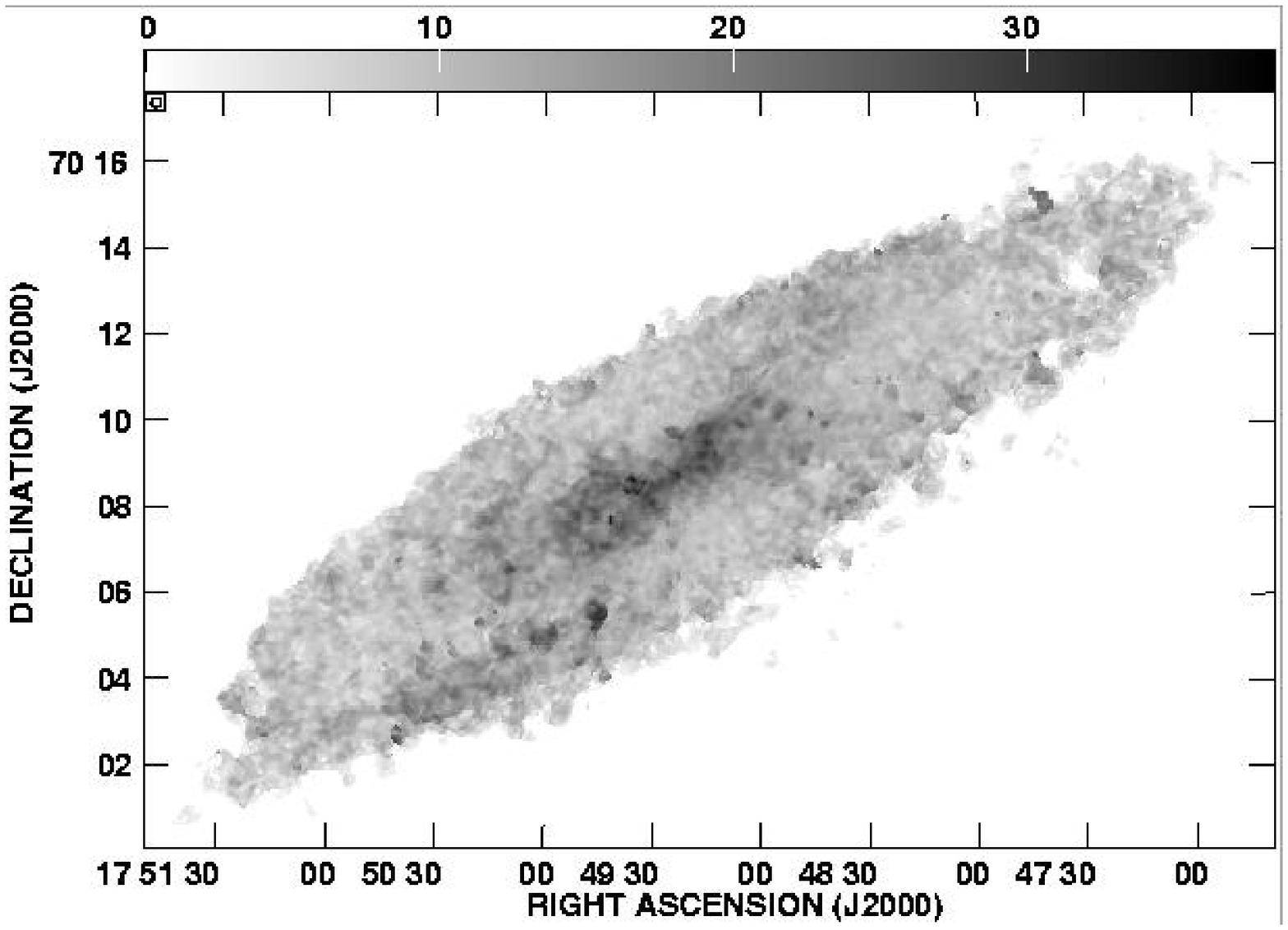}}
\caption{Second moment of the HI distribution in NGC~6503 plotted
  with a linear gray-scale from 0 to 39 km s$^{-1}$.  No correction
  for projection effects has been made.}
\label{fig:Xmom2}
\end{figure}

\begin{figure}
\centering
\resizebox{3.0in}{!}{\includegraphics{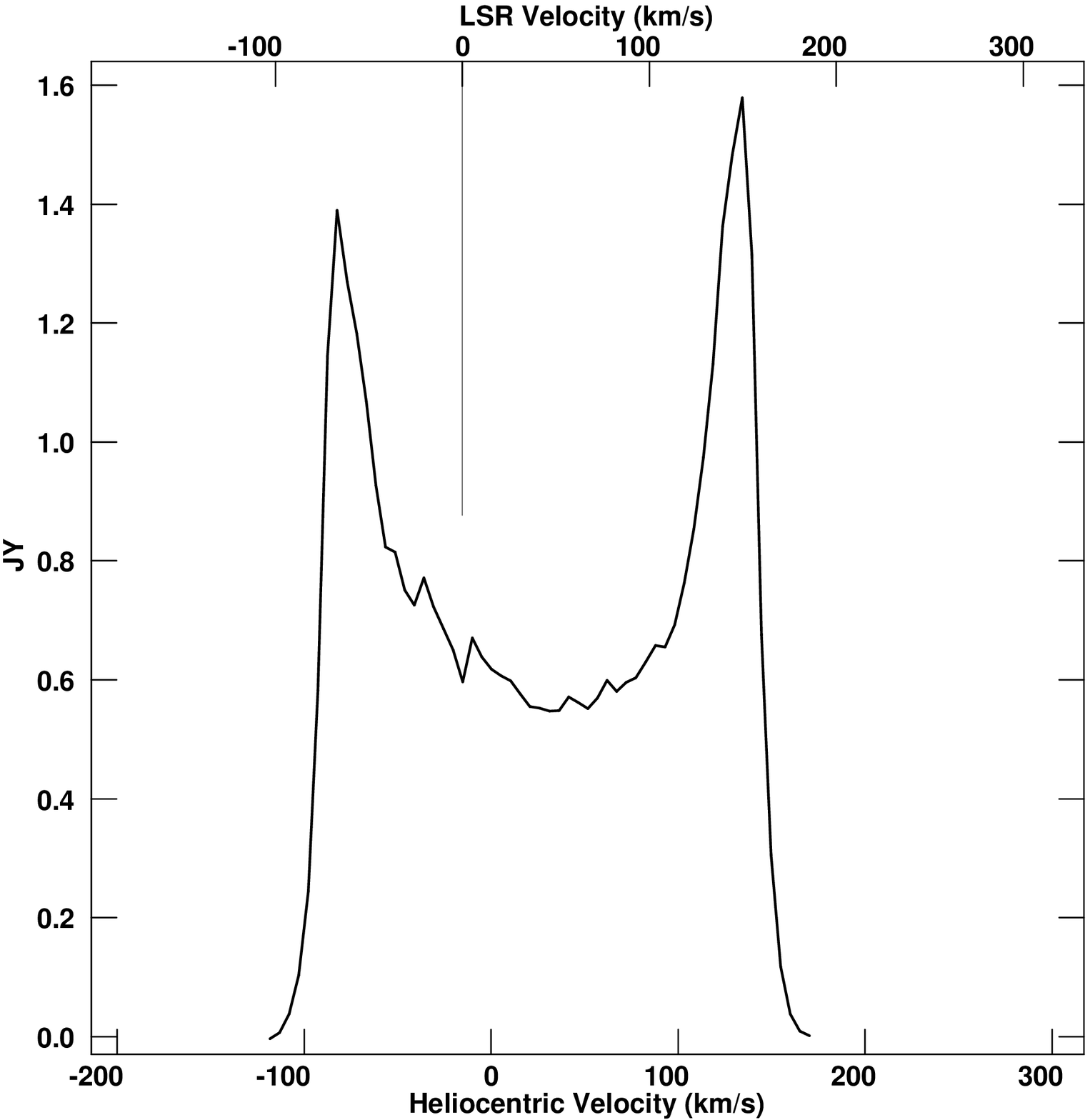}}
\caption{The integrated spectrum of NGC~6503 after primary beam
  correction and blanking of line-free regions.  The LSR as well as
  heliocentric velocities are shown, with a long tick mark at 0 km
  s$^{-1}$ LSR\@.  Errors in the spectrum due to HI emission from
  the Milky Way are evident near this velocity.}
\label{fig:Ispec}
\end{figure}

The blanked image cube may be analyzed along the velocity axis for its
zeroth, first, and second moments.  The zeroth is a measure of total
HI content while the first, if the spectra are narrow and symmetric as
they are in NGC~6503, is a measure of the local velocity.  Under the
same conditions, the second moment reflects the velocity dispersion of
the gas.  The first two moments are shown in Fig.~\ref{fig:Xmoms},
while the second moment is shown in Fig.~\ref{fig:Xmom2}.  In this
section, we focus on the features of the zeroth moment map.  The HI
properties derived from this distribution are summarized in
Table~\ref{t:results}.

The weakest pixels in the zeroth moment image in Fig.~\ref{fig:Xmoms}
have values as low as 1 Jy/beam m s$^{-1}$, but the spectra at these
pixels do not contain convincingly real line signals even in the
smoothed image cube.  Convincing line signals are seen at pixels with
zeroth moments of 3 Jy/beam m s$^{-1}$, yielding the lowest reliable
HI detection from NGC~6503 of $1.8 \times 10^{19} \,\, {\rm cm}^{-2}$.

Integrating over the zeroth moment image yields an integrated flux
$S_{\rm int}  =  205 \pm 1$ Jy km s$^{-1}$.  The corresponding
HI mass is given by
\begin{equation}
  M_{\rm HI} = (2.356 \times 10^5) S_{\rm int} D^2 \,\, {\rm M}_\odot
   \,\,\,,
\end{equation}
where $S_{\rm int}$ is in Jy km s$^{-1}$ and $D$ is the distance to
NGC~6503 in Mpc.  The total HI mass of NGC~6503 is therefore $M_{\rm
HI} = 1.3 \pm 0.2 \times 10^9\,{\rm M}_\odot$, including a
10\%\ uncertainty in $D$.   The integrated spectrum of NGC~6503,
obtained by summing the flux in each channel of the blanked image
cube, is shown in Fig.~\ref{fig:Ispec} and yields the same HI mass
within the uncertainties.  The width of the integrated profile is $W
\sin i = 234$ km s$^{-1}$ at the half-peak points (246 km s$^{-1}$
at 20\%\ points).

Several features should be noted in the zeroth moment image in
Fig.~\ref{fig:Xmoms}.  The first is the appearance of spiral
structure, particularly in the outer parts of the disk.  The
approaching arm along the southeastern (SE) edge is particularly
prominent.  Other arms are visible in the interior to the SE, NE, and
NW\@.  The spiral arms also appear conspicuously in residual images
after fits to a smooth HI distribution by CUBIT
(Section~\ref{analysis:CUBIT} and Fig.~\ref{fig:CubitDiff}).  Note
that, if the spiral arms are trailing arms in the usual sense, the
southern edge of NGC~6503 is closer than the northern edge.

There are also a number of holes in the HI distribution in
Fig.~\ref{fig:Xmoms}, where a distinct minimum is surrounded by a
ridge of HI.  One of these is at the very center of the NGC~6503 with
an apparent FWHM of 20\arcsec.  Deconvolved from the beam, this
diameter is about 14\arcsec\ or 0.35 kpc.  This hole is visible in the
channel and transposed images of Figs.~\ref{fig:Line4} through
\ref{fig:MinAx}.  Presumably either the neutral hydrogen has been
converted to molecular form or it has been ionized or evacuated by the
activity at the center of the galaxy which gives rise to the weak
continuum radio source.  South of the center there is another clear
depression with an apparent width of 17\arcsec, or a deconvolved width
around 0.24 kpc.  Such HI bubbles are seen in other nearby galaxies:
for example, detailed observations of M33 reveal a multitude of these
holes \citep{TBW02} which appear to be correlated with large OB
associations \citep{DH90}.

\citet{SWC81} found that NGC~6503 is considerably more extended to the
SE than to the NW, with 1.25 times more HI on the E or approaching
half.  Measuring the zeroth moment image in some detail, we do not
confirm these early results.  Instead we find that the average HI in
the approaching half is about 11\%\ more per pixel but over 8\%\ fewer
pixels.  The net is only 4\%\ more HI in the approaching side of the
galaxy.  The outer major axis radius to the NW is 900\arcsec\ (22.6
kpc), while it is only 825\arcsec\ (17.2 kpc) to the SE\@.

The mean HI/optical radius ratio is therefore 3.7 for NGC 6503\@.
While this value is larger than typically observed in spirals
\citep{BR97}, it does not rival the relative HI extents of some
low-mass systems (e.g. \citet{Getal07}).

\subsection{Kinematics}
\label{analysis:kin}

The first moment of the HI distribution in Fig.~\ref{fig:Xmoms} shows
that the kinematics of NGC~6503 are highly regular. We find some
evidence for a mild warp in the outer disk, first identified by
\citet{SWC81}: the isovelocity contours in Fig.~\ref{fig:Xmoms} make a
somewhat greater angle with the major axis in the NE and SW quadrants
compared to the NW and SE ones. The emission in some individual
channels of the image cube exhibits the same behavior, notably at
-10.1 km s$^{-1}$ and 67.2 km s$^{-1}$ in Fig.~\ref{fig:Line4}.

The second moment image (Fig.~\ref{fig:Xmom2}) has larger values
toward the center of the galaxy where the steepness of the rotation
curve produces an increase in the line width due to the finite beam
width.  There is also a quite noticeable peak in the second moment
near where the SE arm bends abruptly.  That feature is due to the
presence of a second spectral feature in that area, slightly offset in
velocity from the main feature; see the emission at -54.4 km s$^{-1}$,
-41.0 km s$^{-1}$ and -25.5 km s$^{-1}$ in Fig.~\ref{fig:Line4}.  It
is plausible that the bend in the isovelocity contours in the first
moment at that location (Fig.~\ref{fig:Xmoms}) also stems from this
limited spectral feature.

\begin{figure}
\centering
\resizebox{3.0in}{!}{\includegraphics{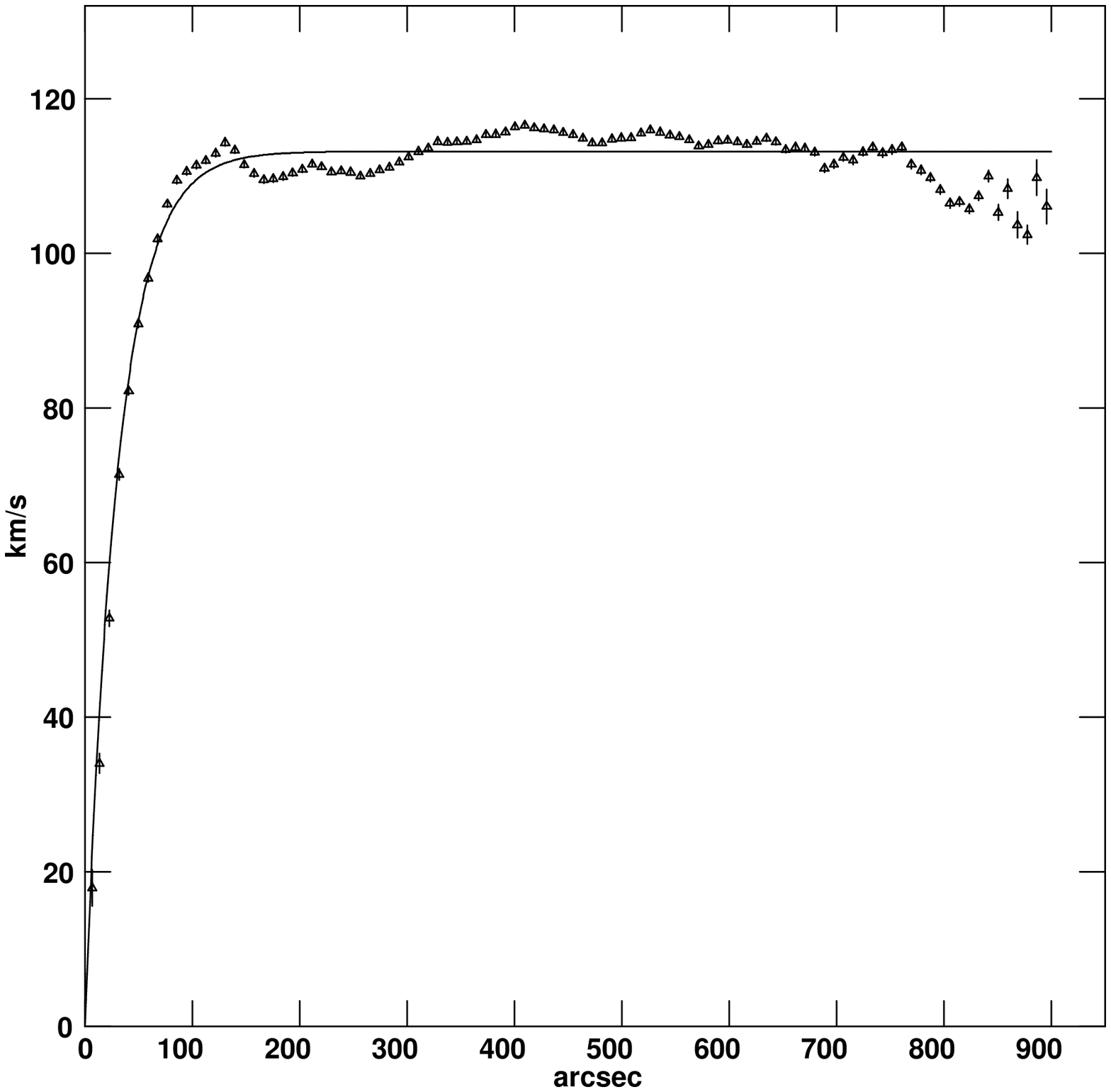}}
\caption{Rotation curve for the HI disk of NGC~6503, obtained from
    the AIPS program GAL by fitting all points in the first moment
    image.}
\label{fig:rc}
\end{figure}

\begin{figure}
\centering
\resizebox{3.0in}{!}{\includegraphics{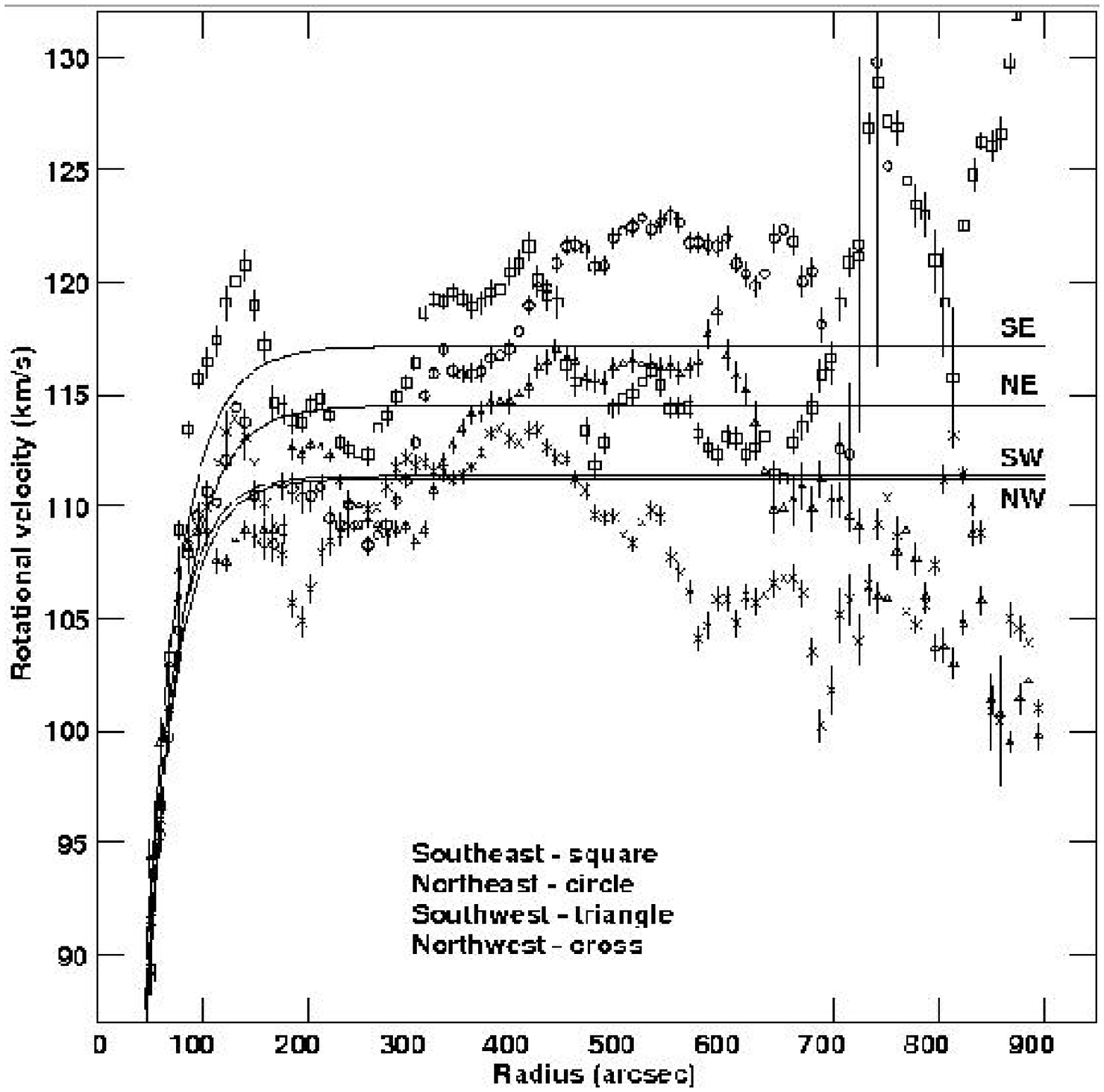}}
\caption{The rotation curve fit by the GAL program for the NW (cross
  symbols), SW (triangles), NE (circles), and SE (squares) quadrants.
  Note that the velocity tick marks are separated by about one
  spectral channel in this blown-up display.}
\label{fig:GALqb}
\end{figure}

\begin{table}
\centering
\renewcommand{\arraystretch}{1.1}
\caption{GAL fit parameters by quadrant \label{t:GALfit}}
\begin{tabular}{cccc}
\noalign{\vspace{2pt}}
\tableline\tableline
quadrant&inclination&$V_{\rm max}$&$R_{\rm max}$\\
 & deg & km s$^{-1}$ & arcsec \\
\tableline
all & 72.6 & 113.2 & 139.2 \\
SW  & 71.7 & 111.4 & 129.7 \\
NW  & 74.3 & 111.3 & 135.5 \\
NE  & 72.7 & 114.5 & 149.9 \\
SE  & 72.3 & 117.1 & 145.7 \\
\tableline
\end{tabular}
\end{table}

Van Moorsel \&\ Wells (1985) describe an AIPS program GAL which fits
rotation curves to user-selected portions of the first moment image
with optional weighting by the zeroth moment image.  We first used
GAL to model the kinematics of the NGC~6503 HI disk as a whole: after
fitting for the kinematic center of the disk, this value was held
fixed and the position angle of the receding major axis and the
systemic velocity were derived.  The best fitting values for these
parameters are given in Table~\ref{t:results}.  Holding these
parameters fixed, we then proceeded to fit for the inclination and the
rotation curve of the system  using the functional form:
\begin{equation}
  V = V_{\rm max} \left( 1 - e^{- \ln(100) R / R_{\rm max}} \right)
\end{equation}
where $V_{\rm max}$ is the maximum velocity of the rotation curve,
which occurs at a radius $R_{\rm max}$.

The rotation curve obtained by fitting all points in the first moment
image is given in Fig.~\ref{fig:rc}, and the extracted parameters are
given in the first row of Table~\ref{t:GALfit}.  The rotation curve in
Fig.~\ref{fig:rc} rises out to $R = 140\arcsec$ (3.5 kpc), and then
remains flat out to $R = 770\arcsec$ (19.4 kpc).  There is some
evidence for a decrease in rotation amplitude beyond this, out to the
last measured point at $R = 900\arcsec$ (22.6 kpc). Given a rotation
curve, the dynamical mass enclosed within a radius $R$ can be computed
assuming spherical symmetry:
\begin{equation}
M_{\rm T} = ( 2.325 \times 10^5 ) R \, V_{\rm rot}^2 \,\, M_\odot \,\,\,,
\end{equation}
where $R$ is in kpc, and $V_{\rm rot}$ is the rotation curve amplitude
in km s$^{-1}$ measured at $R$.  Dynamical mass estimates at different
$R$ for the rotation curve in Fig.~\ref{fig:rc} are given in
Table~\ref{t:results}.

We then extracted the rotation curves and inclinations of NGC~6503 in
the NW, NE, SW and SE quadrants of the disk defined by the kinematic
major and minor axes.  The results from these fits are shown in
Fig.~\ref{fig:GALqb} and Table~\ref{t:GALfit}.  Despite the regularity
of the first moment image for NGC~6503, the discrepancies between the
best fitting parameters for each quadrant suggests the presence of
non-circular motions and geometric distortions in the disk.  The SW
quadrant inclination differs from the NW by 2.6 degrees in the
direction of the plane bending down in that quadrant.  The eastern
side, however, gives an intermediate inclination which is the same in
both quadrants, failing to confirm the second half of the warp
suggested above.  We note that the velocities and physical dimensions 
in the eastern half are somewhat larger than in the western half.
Fig.~\ref{fig:GALqb} makes it clear that there are small-scale
deviations, of order plus/minus one spectral channel (5 km s$^{-1}$),
from simple rotation in all four quadrants.   The deviations are
greatest in the SE and NW in which the view is along spiral arms
against the direction of motion in the SE and in the direction of
motion in the NW\@.  Peculiar velocities due to spiral structure would
be expected in those locations.

\begin{table*}
\centering
\renewcommand{\arraystretch}{1.1}
\caption{Fit parameters from CUBIT \label{t:CUBITfit}}
\begin{tabular}{llccc}
\noalign{\vspace{2pt}}
\tableline\tableline
parameter & units & full galaxy & minimum & maximum\\
\tableline
Position angle         & degrees   & -60.3 & -62.4 & -58.2 \\
Inclination            & degrees   &  75.1 &  73.5 &  76.0 \\
$V_B$                  & km s$^{-1}$ & 118.7 & 118.3 & 119.9 \\
$R_B$                  & arc sec   & 172.6 & 168.5 & 218.7 \\
Brandt index           &           & 0.091 & 0.060 &  0.69 \\
Maximum density        & cm$^{-3}$  & 0.147 & 0.134 & 0.180 \\
Radial density scale\tablenotemark{a} & arcsec    & 353.5 & 329.2 & 373.3 \\
Vertical density scale\tablenotemark{a} & arcsec    &  26.0 &  23.6 &  33.0 \\
Vertical column density & $10^{20} {\rm cm}^{-2}$ & 4.1 & 4.1 & 4.8 \\
\tableline
\end{tabular}
\tablenotetext{a}{the radius or height at which the density falls by a
  factor of $1/e = 0.3679$.}
\end{table*}

\begin{figure*}
\centering
\resizebox{5.in}{!}{\includegraphics{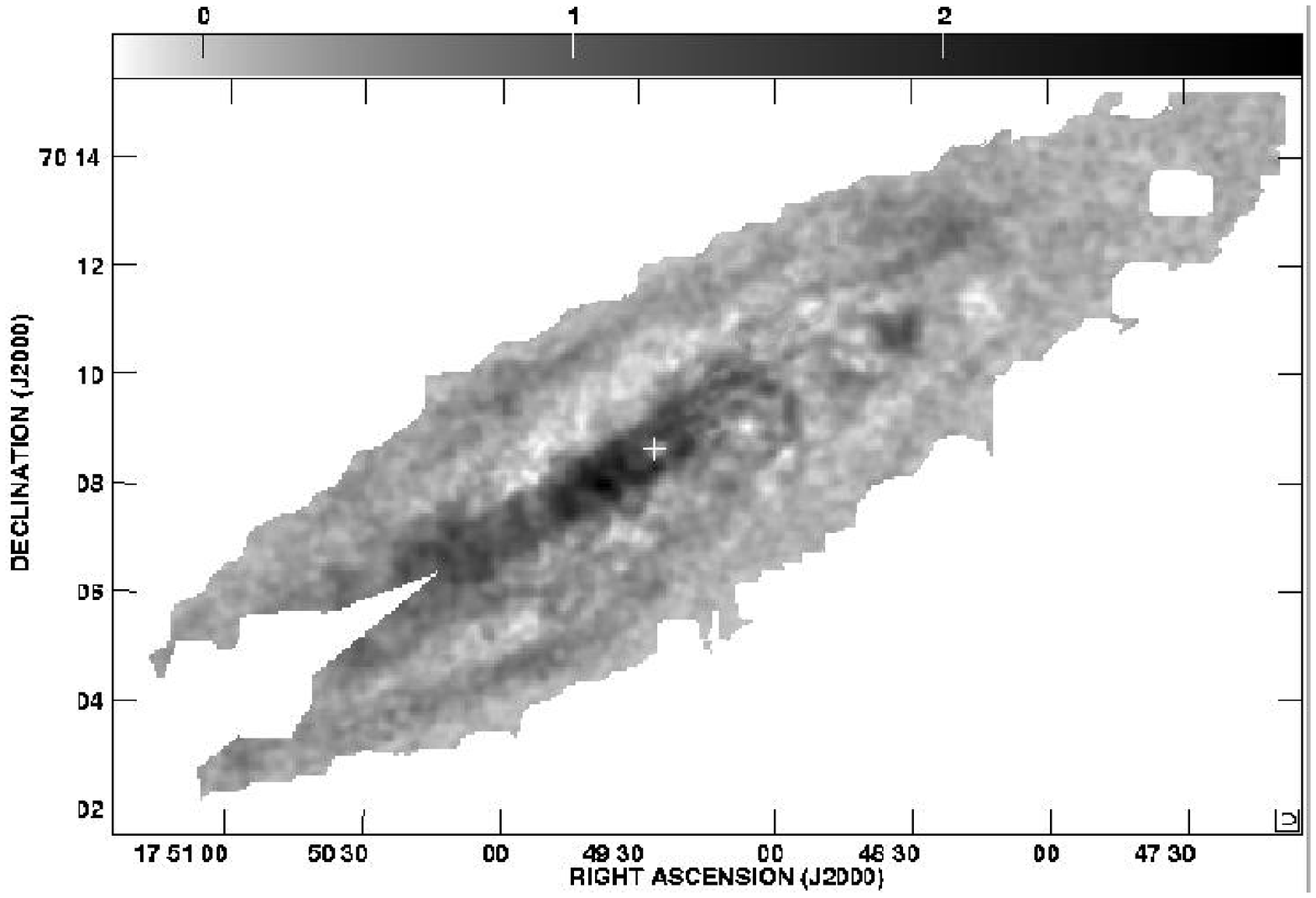}}
\vspace{12pt}
\resizebox{5.in}{!}{\includegraphics{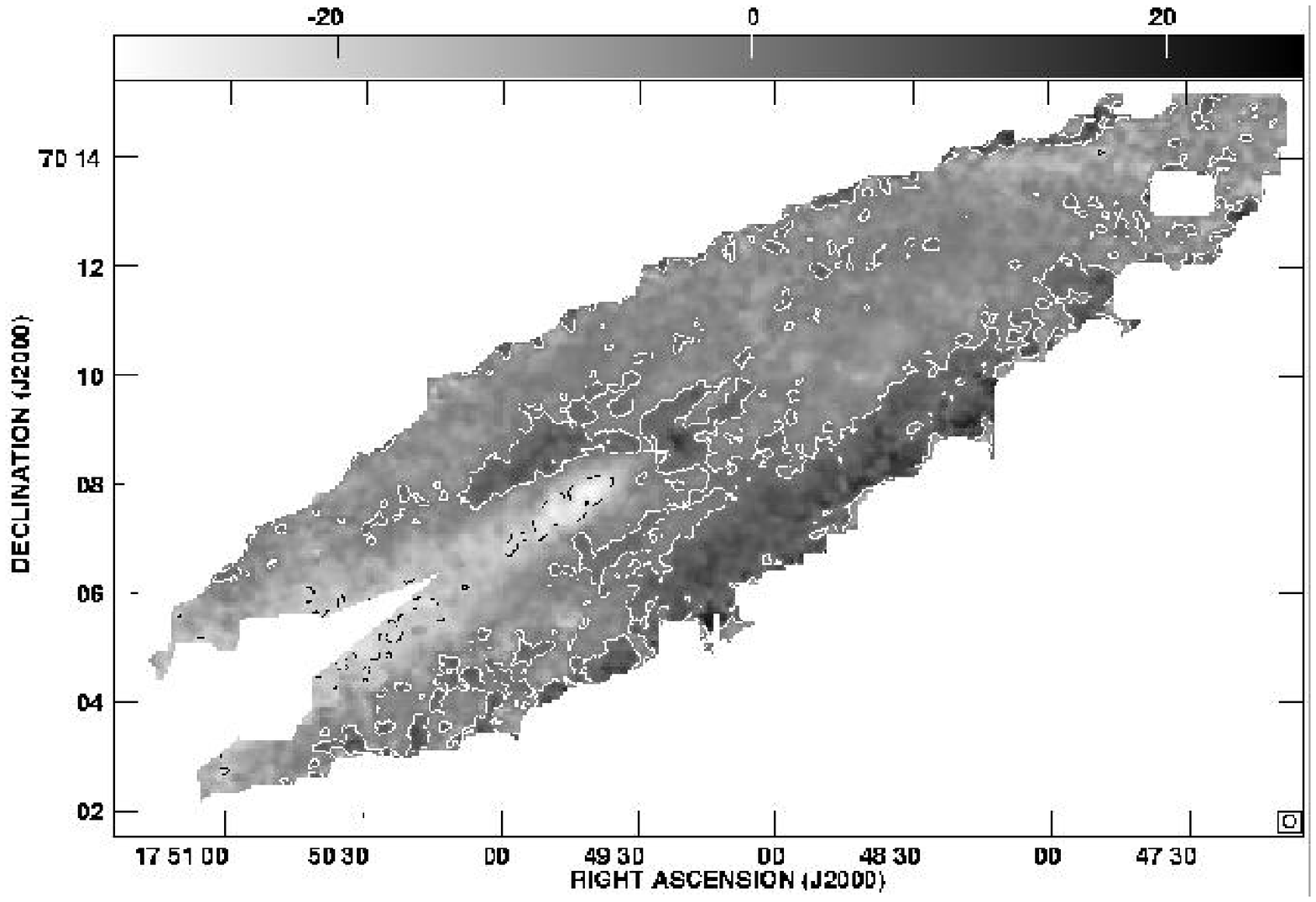}}
\caption{Top: zeroth moment image of NGC~6503 minus that fit by CUBIT
  plotted with a logarithmic gray-scale.  The step wedge is marked in
  steps of $10^{21}\, {\rm cm}^{-2}$.  A comparison of the CUBIT model
  and the radial column density along the major axis is shown in
  Fig.~\ref{fig:RadSD}.  Bottom: first moment image of NGC~6503 minus
  that fit by CUBIT.  The white contour is drawn at 0 km s$^{-1}$, and
  the black, dashed contour is drawn at -20 km s$^{-1}$.}
\label{fig:CubitDiff}
\end{figure*}

\subsection{CUBIT models}
\label{analysis:CUBIT}

Another AIPS task, CUBIT, was initially developed by \citet{I94} to
fit the entire data cube rather than just the first moment image.  The
model assumes that the galaxy is a disk of finite thickness in
circular rotation, and fits for one of several functional forms of the
HI density distribution and rotation curve.  The CUBIT model fit to
NGC~6503 held the kinematic center and systemic velocity fixed to the
values in Table~\ref{t:results} while fitting for a Gaussian density
profile within the plane of the disk, an exponential density profile
perpendicular to the plane, and a Brandt rotation curve given by
\begin{equation}
  V(R) = \frac{V_{\rm B}\, R / R_{\rm B}}{\left(\,1/3\, +\, 2/3
  (R/R_{\rm B})^m\, \right)^{3/(2m)}} \,\,\,
\end{equation}
where $V_B$ and $R_B$ characterize the rotation curve shape and $m$ is
the ``Brandt index.''  The results of the CUBIT fits to the NGC~6503
cube are summarized in Table~\ref{t:CUBITfit}, where the columns
labeled ``minimum" and ``maximum" illustrate the range in parameter
values found when CUBIT is restricted to different regions of the
disk. In general, there is good agreement between the best fitting
parameters found by GAL (Table~\ref{t:GALfit}) and by CUBIT
(Table~\ref{t:CUBITfit}). We note that a higher inclination is
expected from CUBIT, since GAL assumes a thin disk while CUBIT takes
its finite thickness into account.

Fig.~\ref{fig:CubitDiff} shows the residuals when the best fitting
CUBIT model for the disk as a whole is subtracted from the data. The
residuals emphasize the geometric and kinematic deviations from an
ideal disk that were discussed in Sections \ref{analysis:totHI} and
\ref{analysis:kin}.   They are largest along the SE major axis, where
the CUBIT model greatly underestimates the zeroth moment and does not
fit the velocity profile well.  The kinematic discrepancy between the
data and the model might be expected given the rotation curve fits in
Fig.~\ref{fig:GALqb}, where the largest disparity between the
quadrants within the optical radius is between the SE and SW at $R
\sim 130\arcsec$.  The first moment velocities in the data also exceed
the model by up to 20 km s$^{-1}$ along the southern edge.  This
provides some support for the previous suggestion of a warp in the SW.

\begin{figure}
\centering
\resizebox{3.0in}{!}{\includegraphics{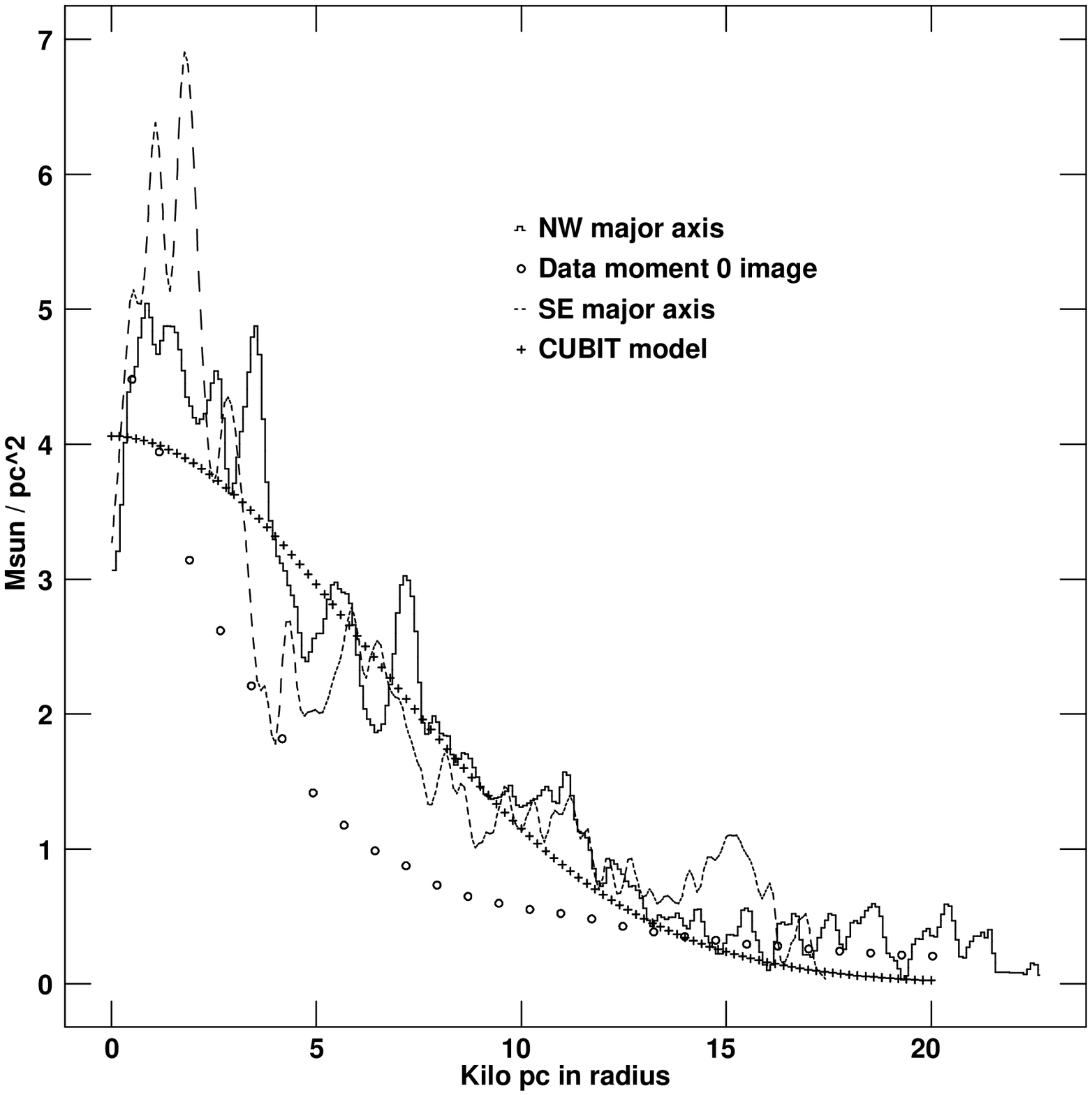}}
\caption{The radial column density in ${\rm M}_\odot / {\rm pc}^2$
  plotted against radius in NGC~6503.  The plus signs trace the CUBIT
  model, the squares are averages over concentric rings in the zeroth
  moment image, and the dashed and solid lines are slices in the
  zeroth moment image along the major axis to the SE and NW,
  respectively.  The slices have been corrected to radial column
  densities by multiplying by the cosine of the inclination
  ($75.1^{\circ}$).}
\label{fig:RadSD}
\end{figure}

Integrating the exponential vertical density profile, the CUBIT model
provides the best fit Gaussian to the radial column density profile in
NGC~6503.  This profile can also be estimated directly from the zeroth
moment image.  Different inclination-corrected measures of the radial
column density in NGC~6503 are shown in Fig.~\ref{fig:RadSD}, where
the plus signs show the CUBIT model output, the squares show the
profile obtained by averaging concentric rings in the zeroth moment
image, and the dashed and solid lines show the profiles obtained from
slices of the zeroth moment image along the major axis to the SE and
NW, respectively.  A comparison between the CUBIT model output and the
SE and NW major axis profiles explains the large zeroth moment
residuals in Fig.~\ref{fig:CubitDiff}: the surface density profile is
well-approximated by a Gaussian along the NW major axis, but deviates
from this functional form by almost a factor of two along the SE major
axis at $R \sim 2\,$kpc (80\arcsec).  As a result, the zeroth moment
residuals in Fig.~\ref{fig:CubitDiff} contain nearly 45\%\ of the
total HI in NGC~6503.  We note that the radial column density profile
estimated from the zeroth moment image is contaminated by the
relatively poor resolution along the minor axis due the high
inclination of NGC~6503: this biases the profile low at intermediate
$R$ and high at large $R$\@.

The peak column densities and other parameters found here are similar
to those found by \citet{vdHetal93} for a number of low surface
brightness galaxies.  Like them, NGC~6503 falls below the average for
field Scd galaxies (8.5 ${\rm M}_\odot\, /\, {\rm pc}^2$) and would
fall below the \citet{K89} criterion for normal star formation.

\subsection{Extra-planar HI Models}
\label{beard}

In comparison to the HI distributions of several nearby spirals
\Citep{FMSO02,MW03,BOFHS05,OFS07}, position-velocity slices along the
major (Fig.~\ref{fig:MajAx}) and minor (Fig.~\ref{fig:MinAx}) axes of
NGC~6503 show little evidence for low-velocity gas and a high degree
of symmetry, respectively. Our CUBIT models of NGC~6503
(Section~\ref{analysis:CUBIT}) suggest that the mean vertical density
scale\footnote{height over which the density drops by a factor of
$1/e = 0.3679$} of its HI layer is 26\arcsec\ (0.65 kpc). Taken
together, the position-velocity slices and CUBIT models thus suggest
that NGC~6503 lacks the thick, lagging extra-planar HI layer that is
commonly detected in spiral galaxy observations at this sensitivity
(see Section 1 for references). In this section, we model the vertical
extent and kinematics of the HI layer in NGC~6503 to place
quantitative constraints on the properties of its extra-planar gas. 
 
To emphasize any large-scale, low column-density features that might
be present in the data, we smoothed the HI cubes presented in
Section~\ref{images:HI} spatially and spectrally to a resolution of 
30\arcsec\ and $10.3\,$km s$^{-1}$.  We compare our models to this
smoothed cube.

We use the GIPSY \citep{vdHetal92} task GALMOD to model NGC~6503 as
one or two HI layers of finite thickness in circular rotation, using 
the results from the analysis in Sections~\ref{analysis:kin}~and~
\ref{analysis:CUBIT} as inputs.  In all of the models, we adopt a
kinematic center, position angle and systemic velocity identical to
the values in Table~\ref{t:results} and a constant velocity dispersion
of 8 km s$^{-1}$ (Fig.~\ref{fig:Xmom2}).        
 
Our CUBIT models suggest that the receding (W) side of the disk is
better approximated by a smooth HI layer than the approaching (E) side
(Fig.~\ref{fig:CubitDiff}), and we therefore focus on reproducing its
properties with GALMOD\@.  Specifically, the input model kinematics are
the rotation curve extracted by GAL for the receding half of the disk,
and the input HI surface mass density is that along the NW major axis
(Fig.~\ref{fig:RadSD}).  We have corrected the input rotation curve and
HI surface density for beam smearing at $R< 100\arcsec$.  The output
models are smoothed to the same spatial and spectral resolution as the
smoothed data cube.

\begin{figure*}
\centering
\resizebox{5.13in}{!}{\includegraphics{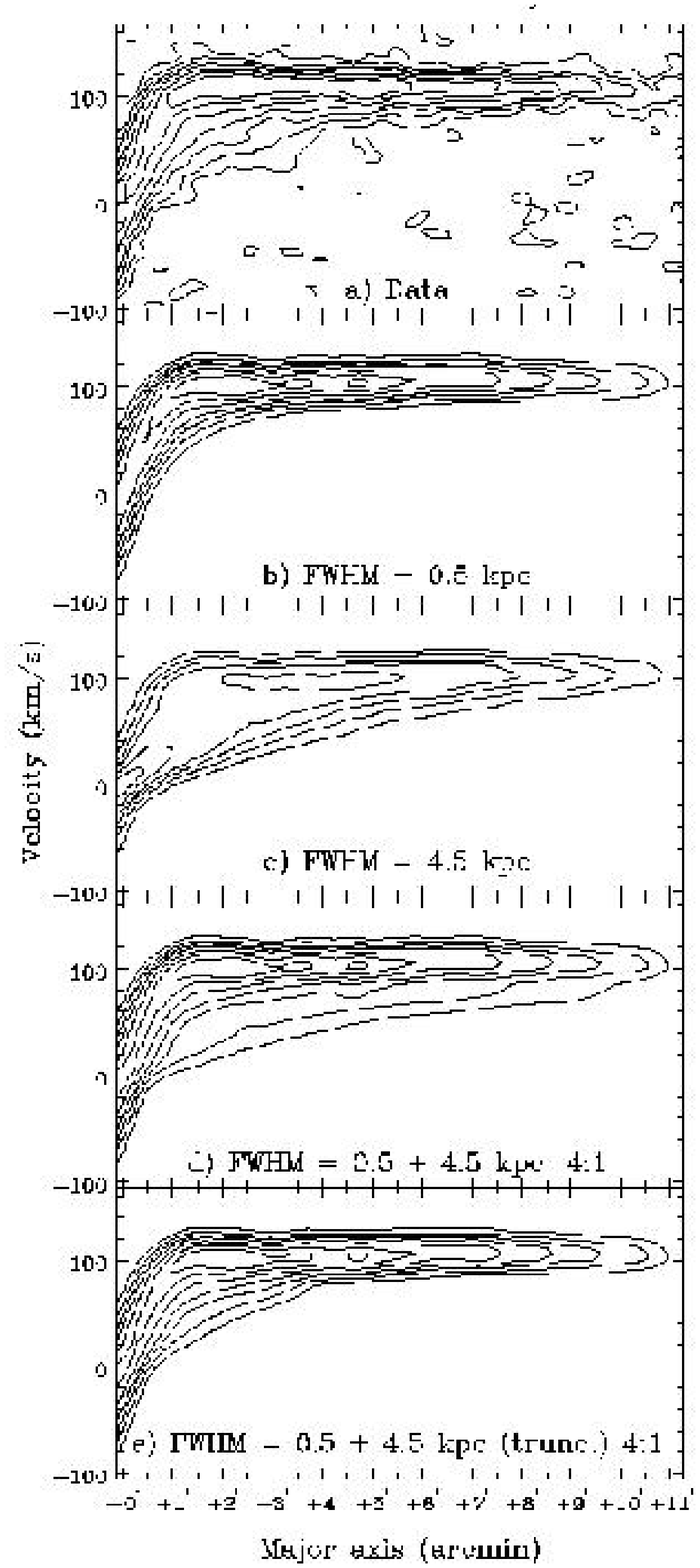}\ %
    \includegraphics{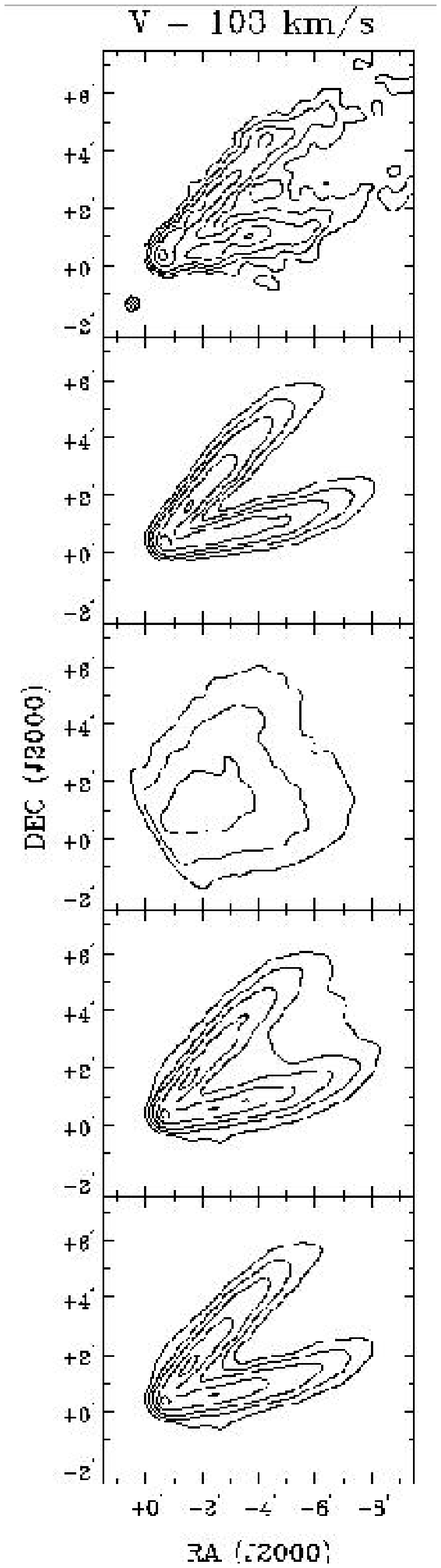}}
\caption{Position-velocity slices along the NW major axis (left)
  and channel map at $V = 103.3\,$km s$^{-1}$ (right) in a) the
  smoothed image cube, and b) -- e) various GALMOD models. The
  contours in the position-velocity panels are at $
  (1,\,3,\,6\,,10\,,18\,,28\,,40) \times 0.76$ mJy/beam, and the
  contours in the channel map panels are at 1.5, 4 ,8 ,16 and 24
  mJy/beam. Panels b) -- e) show our most plausible models including
  b) a single thin HI layer, c) a single thick HI layer, d) thin and
  thick layers with a column density contrast of 4:1 where the thick
  disk lags the thin one by $25\,$km s$^{-1}$, e) thin and thick
  layers with a column density contrast of 4:1 where the thick disk
  lags the thin one by $25\,$km s$^{-1}$ and is truncated at the
  optical radius ($R = 3.55\arcmin$).}
\label{fig:galmod}
\end{figure*}
 
A selection of galaxy models is compared to the smoothed NGC~6503
image cube in Fig.~\ref{fig:galmod}.  The left column shows a
position-velocity slice along the NW major axis, and the right column
shows the detected HI emission in the channel at $V = 103.3\,$km
s$^{-1}$.  The contours in all rows of a column are identical.  We
choose to show these model properties because we find that they
provide the best diagnostics of the vertical extent and kinematics of
the HI layer.  We note that because of projection effects there
was little difference between the position-velocity slices along the
minor axis for various models, and they are not included in
Fig.~\ref{fig:galmod}.
  
We first model the HI in NGC~6503 as a single Gaussian layer.  
Figs.~\ref{fig:galmod}b and \ref{fig:galmod}c show the model
morphologies for HI layers with FWHM = $0.5\,$kpc and FWHM =
$4.5\,$kpc, respectively.  They illustrate that a model with a thin HI
layer (Fig.~\ref{fig:galmod}b) reproduces the gross features of the
major axis position-velocity slice beyond the optical radius ($R =
3.55\arcmin$) as well as the observed channel maps: this is expected
given the low vertical density scale returned by CUBIT
(Section~\ref{analysis:CUBIT}).  However, this model is unable to
reproduce the position-velocity contours within the optical disk:
there is more anomalously low-velocity gas at these radii than
expected. Fig.~\ref{fig:galmod}c demonstrates that invoking a thick HI
layer does produce low-velocity gas at small $R$ in the
position-velocity slice, but that both the model channel map and
position-velocity morphologies are inconsistent with the data.
 
We thus proceed to model NGC~6503 with both a thin and a thick layer.
As in \citet{SSS00}, we find that plausible models with co-rotating
thick and thin layers have similar morphologies to our one-component
models.   We therefore explore models where the thin and thick layers
have the same column density profile shape and rotation profile
shape, but where the thick layer has a total column density and
rotation velocity that are lower than that in the thin layer.
 
Fig.~\ref{fig:galmod}d shows one of our better models of NGC~6503
with this morphology.  The thin layer has a FWHM of $0.5\,$kpc.  The
thick layer has a FWHM of $4.5\,$kpc, and rotates $25\,$km s$^{-1}$
slower at all $R$ compared to the thin disk.  The thin:thick layer
column density ratio is 4:1.  We see that the morphology of the
position-velocity slice within the optical disk approaches that seen
in the data.  However, the model also predicts detectable low-velocity
gas at larger $R$ in the position-velocity slices as well as
large-scale, low-intensity emission in the channel maps which are not
observed in the data.  Increasing the column density ratio between the
HI layers alleviates the discrepancies in the outer disk and channel
maps, but the models then fail to reproduce the ``beard" across the
optical disk for plausible differences in velocity between the
layers.  We therefore find that this class of models, too, fails to
reproduce the position-velocity slice and channel map morphology at
all locations in the HI disk.
  
Finally, we invoke 2-component models where the thick, low
column-density, lagging HI layer is truncated beyond the optical
radius.  One of our better models with this morphology is shown in
Fig.~\ref{fig:galmod}e.  The thick layer has a FWHM of $4.5\,$kpc, and
rotates $25\,$km s$^{-1}$ slower at all $R$ compared to the thin disk.
Within the optical radius ($R = 3.55\arcmin$), the thin:thick layer
column density ratio is 4:1, and the thick layer has zero column
density farther out.  This model is the only one in
Fig.\ref{fig:galmod} that reproduces the morphology of the major axis
position-velocity slice as well as the channel maps. While not shown
here, this is also the case for the advancing half of the disk, as
well as for other channels and slices in the cube.
 
Given the sensitivity of our data, we estimate that truncated 2-layer
models with combinations of column density ratios between 4:1 and 6:1,
velocity lags between $20\,$km s$^{-1}$ and $40\,$km s$^{-1}$, and
thick layer FWHM between $3\,$kpc and $5\,$kpc could reproduce the
data. However, the column density ratio between the thin and thick
disks outside the optical disk must be $\gtrsim$10:1 in order to
escape detection at the 3-5$\sigma$ level.  We therefore conclude that
the extra-planar HI layer in NGC~6503 extends only out to its
optical radius.

\section{Discussion and conclusions}

We have presented aperture synthesis observations of the nearby,
late-type spiral galaxy NGC~6503.   Advances in calibration and
imaging software enabled us to produce a wide-field continuum image at
a dynamic range of 9000:1.  We detect HI emission at column densities
as low as $1.8\times 10^{19}\,{\rm cm}^{-2}$, with a spatial
resolution of 0.35 kpc at the galaxy (Fig.~\ref{fig:Xmoms}): our
images thus have 7 and 6 times higher sensitivity and resolution than
those of \citet{vMW85}, respectively.  The HI cube is also to a great
extent free of contamination due to missing short spacings.  In
Appendix~\ref{ap:imagr},  we discuss the improved imaging algorithms
available within the AIPS software package which have made these
images possible.

Even at the higher sensitivity and resolution of our observations
relative to previous studies, we find that the HI disk in NGC~6503
is remarkably regular (Figs.~\ref{fig:Line4},~\ref{fig:Xmoms}).
Nonetheless, several deviations from an idealized distribution are
apparent: the outer portions of many emission ``wings" bend away
from the major axis in the channel maps, and the zeroth moment
reveals a number of density enhancements giving the impression of      
large-scale spiral arms as well as smaller spiral-like structures
closer to the dynamical center. The HI density is reduced at the
dynamical center and at several other locations in poorly-resolved
holes about 0.3 kpc in diameter.

We trace the rotation curve of NGC~6503 out to 3.7 optical radii
(Fig.~\ref{fig:rc}), and find that it is flat almost out to the last
measured point.  Fitting rotation models to the first moment image by
quadrants reveals kinematic features comparable in size to the
channel width (5.15 km s$^{-1}$), some of which probably reflect
spiral density wave disturbances  (Fig.~\ref{fig:GALqb}).  This is
also evident in our CUBIT models of the HI density distribution and
kinematics (Fig.~\ref{fig:CubitDiff}): the velocity field is distorted
from the fit model in the SW portion of NGC~6503, partially confirming
the presence of the warp in the plane suggested by \citet{SWC81}.
However, a comparable disturbance to the kinematics is not observed on
the opposite side of the plane.  In addition, the HI density is
greatly enhanced compared to, with a significant velocity discrepancy
from, the fit model along the SE major axis.  Careful consideration of
these asymmetries will be taken in forthcoming dynamical models.

Despite the greater HI emission-line sensitivity of some recent
observations of nearby spirals compared to those presented here
\citep[e.g.][]{FMSO02, OFS07}, we find that the disk of NGC~6503
extends to a similar limiting column density (Fig.~\ref{fig:Xmoms}).
In the images of \citet{FMSO02}, the disk of NGC~2403 truncates at
$\sim2\times 10^{19}\, {\rm cm}^{-2}$ (see their fig. 6), and the
lowest contour in the \citet{OFS07} map of NGC~891 is at $1.0 \times
10^{19}\, {\rm cm}^{-2}$ (see their fig. 1; they do not discuss an
actual examination of the spectra at such points).  We trace the HI
disk in NGC~6503 out to $1.8 \times 10^{19}\, {\rm cm}^{-2}$  (22.6
kpc). But the absence of absorption towards the quasar 1748+700 yields
an upper limit of $5\times 10^{17} \, {\rm cm}^{-2}$ for the column
density of cold HI gas along a line of sight which should intersect
the disk at 29 kpc (Fig.~\ref{fig:QuaSpec}). Our results therefore
support the remark by \citet{OFS07}  that significant improvements in
sensitivity have not increased the observed radial size of HI disks.
Perhaps we are observing the cutoff suggested by \citet{SS76} due to
ionization by an intergalactic UV radiation field.

While disks do not ``grow" radially when observed at higher
sensitivity, nearly all exhibit extra-planar HI at lower velocities
than those in the plane of the disk (see Section~\ref{s:intro} for
references).  Our images, when smoothed to achieve greater
sensitivity, are no exception (Fig.~\ref{fig:galmod}).  We model the
vertical extent and kinematics of this extra-planar HI, and find that
the channel maps and position-velocity images of NGC~6503 are best
described by 2-component models with a thin disk and a low column
density, lagging thick disk that is truncated at the optical radius.
We estimate that 2-component models with combinations of thin:thick
layer column density ratios between 4:1 and 6:1, velocity lags between
$20\,$km s$^{-1}$ and $40\,$km s$^{-1}$, and thick layer FHWM between
3 kpc and 5 kpc could reproduce the data. However, the thin:thick
column density ratio beyond the optical radius must be $\gtrsim10:1$
for consistency with our images.

There is therefore a clear correlation between the presence of
extra-planar gas in NGC~6503 and its stellar disk.  This morphology is
hard to reconcile with models in which accreting gas forms the thick
HI disk \citep{BCFS06,KMWSM06}.  NGC~6503 therefore provides clear
evidence for a galactic fountain origin for most of the extra-planar
gas in nearby galaxies, as suggested by the models of NGC~2403 and
NGC~891 by \citet{FB08}.  If 10--20\% of the extra-planar gas does
stem from accretion as suggested by \citet{FB08}, it may be possible
to detect its interaction with the {\it thin} disk beyond the optical
radius of NGC~6503.  Considering our limits on the column density
ratio between the thick and thin layers, observations with 2--3 times
the sensitivity of those presented here would be needed to detect this
phenomenon.  Such sensitive observations would also allow for a
detailed search of High Velocity Cloud analogs in a galaxy with a
modest star formation rate, which may shed some insight into their
origin as well.

\begin{acknowledgements}

The authors would like to thank Miller Goss for his encouragement and
many helpful comments on this manuscript.  EWG also had useful
discussions with Juan M. Uson, Morton S. Roberts, and David E. Hogg.
KS acknowledges support from a research grant from the National
Sciences and Engineering Research Council of Canada.

The National Radio Astronomy Observatory is a facility of the (U.S.)
National Science Foundation operated under cooperative agreement by
Associated Universities, Inc.  

\end{acknowledgements}

\facility{NRAO(VLA)}

\appendix

\section{Image Deconvolution in IMAGR}
\label{ap:imagr}

The imaging and deconvolution of most data collected by the VLA are
performed with the AIPS task IMAGR.  Since a number of algorithms
within this program have a significant impact on the quality of the
resulting images, we elaborate on its functionality here.  We focus on
the multi-scale Clean implementation in IMAGR, and illustrate some of
its advantages over the standard point-source Clean using the NGC~6503
data presented in this paper.  A similar implementation of the
multi-scale Clean is described by \citet{C08} and an extensive study
of its use in imaging HI in galaxies has been described by
\citet{Richetal08}.  Their conclusions regarding this method are
similar to ours.

\subsection{Cotton-Schwab-Clark Clean}
\label{ap:cscclean}

Clean was proposed by \citet{H74} as a method to determine the sky
brightness distribution of an object from incompletely sampled
visibilities.  A review of Clean and its variants can be found in
\citet{CBB99}; briefly, a suite of suitable models are convolved by
the beam synthesized from the visibility data and iteratively
subtracted from an image until a user-specified criterion is reached.
The model components are then convolved with a ``Clean" representation
of the synthesized beam (typically a Gaussian) and added back into the
residual image to form the sky brightness distribution.

When the Clean algorithm works solely in the image domain, the
convolution of the model with the synthesized beam must be carried out
over an area four times larger than that to be Cleaned. \citet{BGC80}
pointed out that a smaller ``beam patch'' could be used to find a
modest number of model components, after which the residual image
could be re-computed by subtracting the current source model fully.
This image-based Clark Clean did the subtraction via Fourier
transforms of images.  Cotton and Schwab \citep[see][]{S84} were the
first to suggest and implement an extension of this method.  In the
``Cotton-Schwab-Clark'' version of Clean, after the modest number of
model components are found, their visibilities are subtracted from the
residual visibility data and a new residual image is formed; the
process then continues.  This allows model components to be found over
most of the image while avoiding Fourier transform aliasing and other
inaccuracies in the model subtraction.  Cumulative computational error
in the imaging is also reduced since the images are continually
recomputed using the current residual data which gradually approaches
the noise levels.  The Cotton-Schwab-Clark Clean is the default
deconvolution algorithm in IMAGR.

Unconstrained, the Clean algorithm can move flux density from real
objects into erroneous locations \citep{NVSS98}.  The synthesized beam
of well-sampled visibility data from the VLA reduces this ``Clean
bias,'' because of the relatively low sidelobe levels.  To reduce it
further, Cleans are typically restricted interactively (``boxed'') to
those areas clearly containing real emission.  Any emission regions
too weak to be included will have negligible sidelobes and, thus, do
not need to be deconvolved. We compare the flux densities measured
from images deconvolved with unrestricted and boxed Cleans for the
particular case of NGC~6503 in Appendix~\ref{ap:6503}.

\subsection{Wide-field imaging}
\label{ap:widefield}

When images are constructed with two-dimensional Fourier transforms
from interferometers that are not co-planar, errors occur which
increase as the square of the angular distance from the phase
reference point.  In order to avoid serious distortions in source
shapes and coordinates, images of large areas must be made up of
rather smaller facets \citep{CP92}.  The choice of facet location is
significant, since aliasing and other numerical errors render the
edges and, especially, the corners, of images unreliable.  AIPS
implements an algorithm due to W. D. Cotton in which the facets
are placed on circles so that the full area may be Cleaned without
using the facet corners. 

When imaging a particular facet, IMAGR rotates the projected baseline
coordinates and the observed phases to the phase reference position at
the facet center.  An inverse rotation is performed when the model
components in the facet are subtracted from the current residual data.
These rotations are simple matrix multiplications and and generally
have a negligible computational cost.  Additionally, the time required
to deconvolve correctly-rendered emission is considerably less than
that required to attempt to deconvolve distorted renderings.  The
computed synthesized beam varies slightly from facet to facet.  IMAGR
differs from other faceting implementations in that it selects one
facet to be Cleaned in the current cycle, finds new model components
in that facet, and subtracts them from the current residual data.
When it selects the next facet to Clean (see
Section~\ref{ap:imagrMS}), it re-images that facet using the latest
residual visibility data.  In this way, sidelobes of a strong source
in one facet are therefore removed from the others before the latter
are Cleaned, which avoids a significant bias in the final image that
is otherwise hard to mitigate.

\subsection{Multi-scale Clean in IMAGR}
\label{ap:imagrMS}

Clean was originally described as a method to model a sky brightness
distribution as a collection of point sources \citep{H74}.  While a
point-source Clean deconvolves poorly-resolved objects accurately and
efficiently, it requires a large number of components to represent
more extended objects and those representations can have non-physical
attributes \citep[e.g.,][]{C83}.  In particular, Clean has a bias
against pixels adjacent to one recently selected.  To mitigate
artifacts in the final image from this bias, model objects are
restored to the residual image only after convolution with a ``Clean
beam.''  Nonetheless, images of extended objects deconvolved with a 
point-source Clean tend to show systematic corrugations that can be
confused for real structure \citep[e.g.,][]{CBB99}.

As early as 1974, one of us (EWG, unpublished) began investigating the
use of Gaussian source models in the Clean algorithm instead of
point-source models.  Extended sources of a single size cannot
accurately model point sources, however, and are of little interest in
an astronomical context since most fields contain unresolved objects.
In order to mitigate these difficulties, \citet{WS88} developed an
image-based algorithm which simultaneously solved for point components
{\it and} Gaussians of a single pre-determined size.  \citet{HC99}
described an extension of this algorithm which would model the image
with Gaussians of a few selected widths as well as with points.  Their
algorithm was an image-only Clean with ``cross beams'' for subtracting
components of one scale from images made for another scale.  At each
iteration, a component was found in the sky brightness distribution
imaged on a particular scale, and then subtracted from the
distributions imaged on all scales.  It was said that this algorithm
could run with high loop gain\footnote{the amount of each component
subtracted at each iteration, usually 0.1, but values close to 1.0
were said to work well} and almost no user-set steering parameters.

The multi-scale Cotton-Schwab-Clark Clean implemented in IMAGR is
rather different from the \citet{HC99} algorithm, and has now been
successfully applied to a number of datasets
\citep[e.g.][]{TBW02,CE06,MRCT06,Ketal08}.  The user specifies the
number of circular Gaussians to be used as model components, as well
as their widths.  Although it is not  required, one of these Gaussians
should have a width of zero to model unresolved objects and other
small-scale structures accurately.  IMAGR converts these widths into
tapers (Gaussian functions which lower the weights at larger projected
baseline lengths) for each scale and uses them to make images of the
beam and the data.  These tapered beam images are used to correct the
data images to units of Jy/beam and to determine the Gaussian Clean
beam for each scale.  IMAGR then discards the beam images and
constructs new beam images at each scale representing the convolution
of the tapered beam with the Clean beam.  This process is carried out
rather simply by re-making each beam image with a taper $\sqrt{2}$
smaller, where the taper parameter in AIPS is the baseline length at
which the tapering Gaussian function reaches 30\%.

IMAGR Cleans each resolution for each facet separately, treating $M$
spatial facets times $N$ scales as $MN$ total facets.  At each major
cycle, it selects one of the $MN$ facets to Clean, images that facet
with the current residual visibility data, finds Clean components in
that facet, and subtracts the Fourier transform of those components
from the residual visibility data.  It stops Cleaning a particular
facet when the peak flux density therein falls below a scale-dependent
cutoff, and the cleaning stops altogether when all facets satisfy this
or another termination criterion.  The components on all scales are
then restored to the image made at the finest (usually point-source)
scale.  These components are added back as Gaussians of the size
fitted to the synthesized beams and scaled by them to produce an image
with correct units.

This algorithm does not work well without some steering.  The user may
specify a variety of multi-scale Clean control parameters in IMAGR,
but only two of them have turned out to be important.  The first is
the resolution-dependent flux density cutoff discussed above, passed
to IMAGR via the parameter FGAUSS\@.  In general, lower resolution
images should have higher flux density cutoffs measured in Jy per the
beam area of that resolution.  The second important steering parameter
controls the order in which facets are Cleaned.  When choosing which
facet to Clean, IMAGR selects the one with the largest weighted peak
flux density $F_w$:
\begin{equation}
F_w = \frac{F}{R^b} \,\,\,,
\label{mscale}
\end{equation}
where $F$ is the peak flux density of the facet, $R$ is the ratio of
the area of the Clean beam for the facet being Cleaned to the area of
the smallest Clean beam, and $b$ is a user-specified ``bias''
parameter that is passed to IMAGR via the parameter IMAGRPRM(11)\@.
The general idea is to force IMAGR to select different scales with
equal probability in the next major cycle.  Thus, if the sky
distribution is entirely composed of point sources, $F$ will be
the same at all scales and the user should set $b = 0$.  In the other
extreme, if the sky distribution is very large and fills the image,
then $F$ in Jy/beam will be proportional to the beam area of the facet
being Cleaned,  and the user should set $b = 1$.   In practice we find
that $0.2 < b < 0.7$ is appropriate for data obtained at the VLA, the
precise value depending on the structure of the sky brightness
distribution.

Multi-scale Clean tends to be faster than point-source Clean despite
the added number of facets.  The lower resolution facets Clean out a
large portion of the total flux density with a modest number of
components, leaving the high-resolution Clean to image point sources
and the edges of objects left behind by the circular Gaussian models.
Because it makes better use of the short-spacing data, multi-scale
Clean also tends to reduce the negative ``bowls'' surrounding
sources observed with missing short-spacing visibility data
\citep[e.g.,][]{CBB99}. This is illustrated in the specific case of
NGC~6503 below.

\subsection{Example: NGC~6503}
\label{ap:6503}

\begin{figure}
\centering
\centerline{\resizebox{3.0in}{!}{\includegraphics{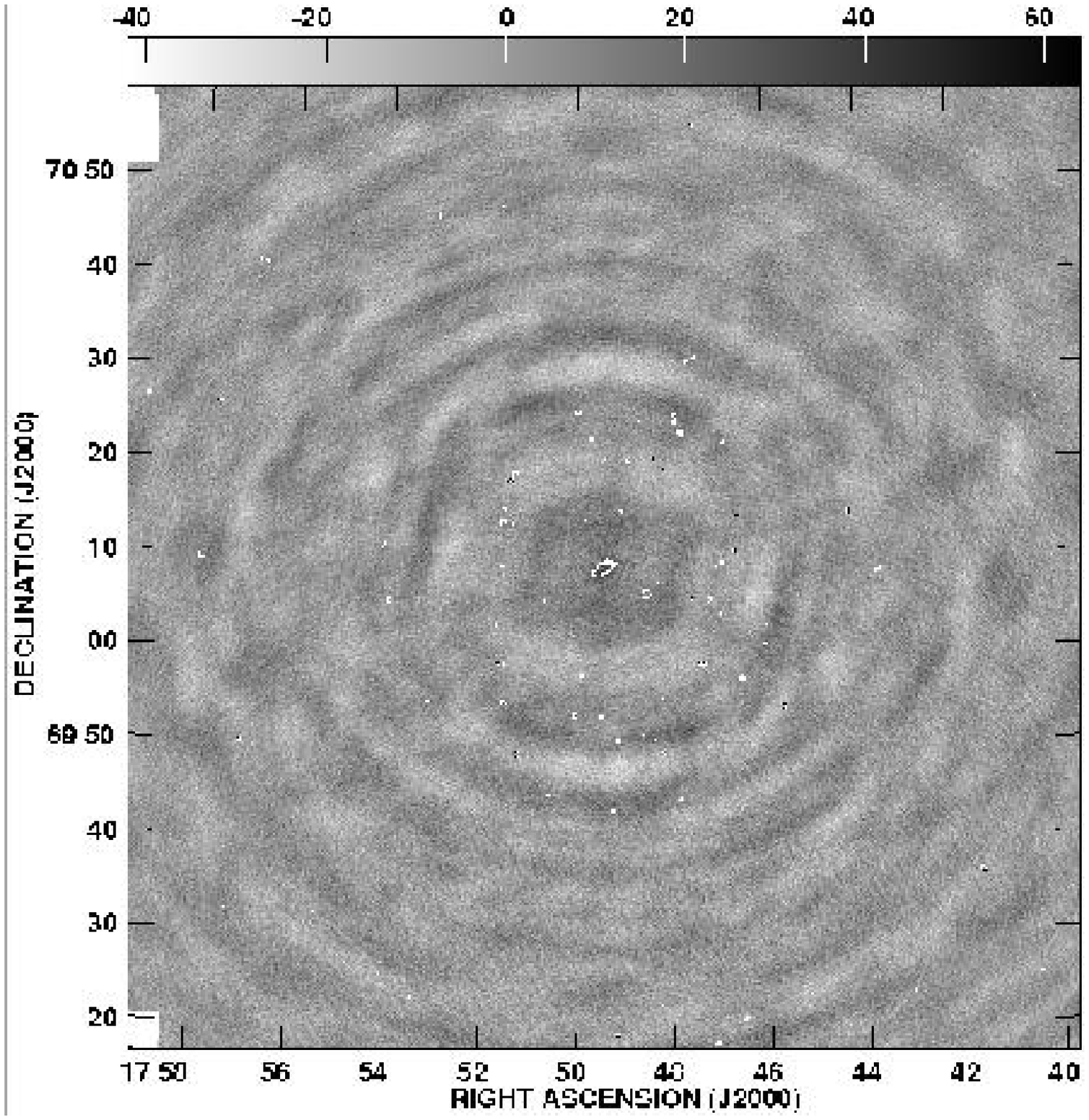}}%
 \hspace{1em}\resizebox{3.0in}{!}{\includegraphics{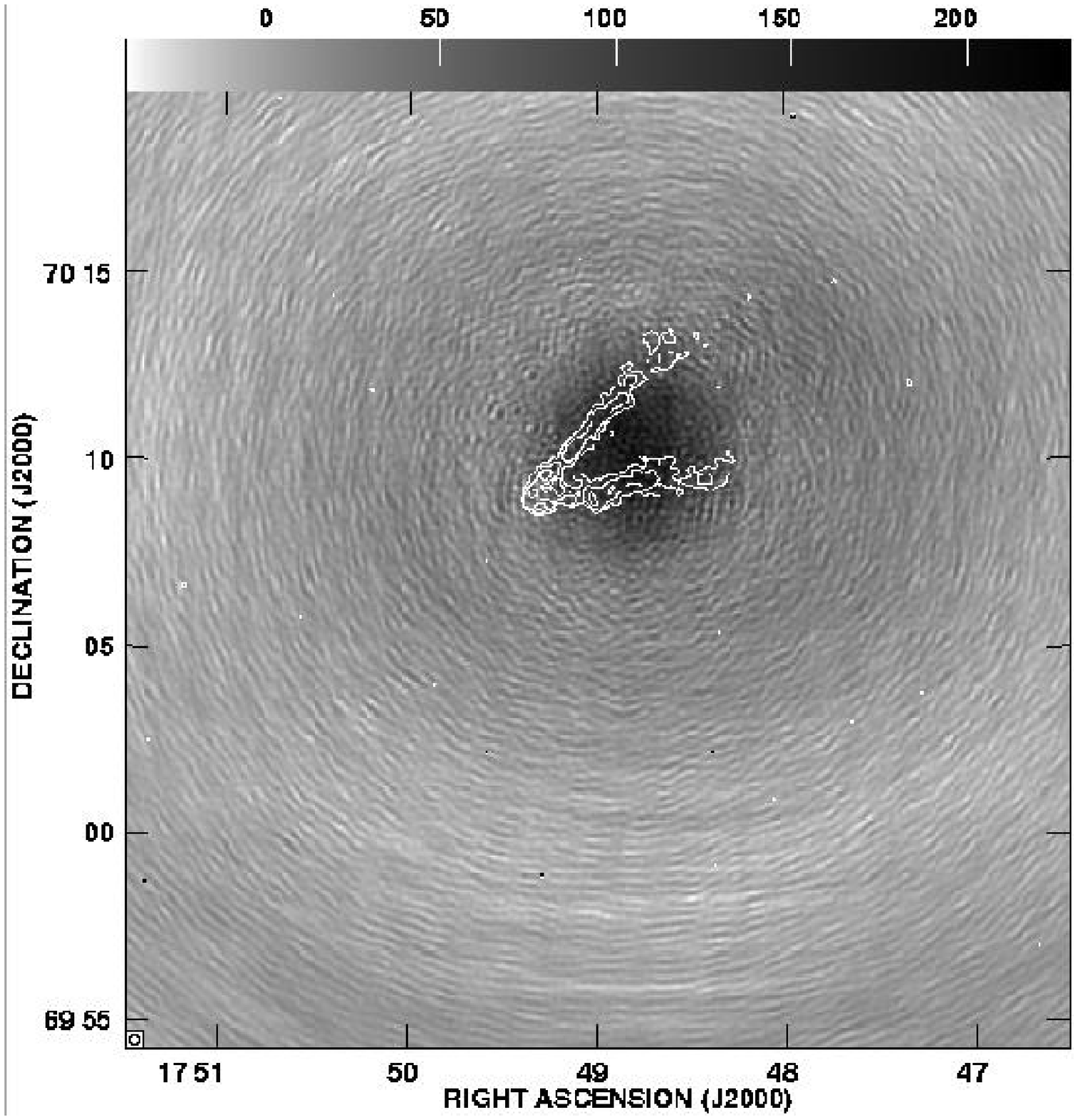}}}
\caption{In gray-scale, the multi-scale Clean image minus
  point-source Clean image of (left) the continuum emission and
  (right) the HI line emission at $V=98.3\,$km s$^{-1}$ in NGC~6503.
  The gray-scale ranges are shown in the step wedges, labeled in
  $\mu$Jy.  The multi-scale images in the continuum and the HI are
  sketched in with contours.  The Clean beam is illustrated at the
  lower left corner in both images and is nearly invisible on these
  scales.}
\label{fig:MSdif}
\end{figure}

The images of NGC~6503 presented in this paper were produced with
IMAGR's multi-scale Clean algorithm (see Section~\ref{S:obs}). Below,
we use these data to illustrate the advantages of the multi-scale
Clean over the default point-source Clean.

The continuum in the vicinity of NGC~6503 was imaged and deconvolved
with a multi-scale Clean over an area 2.5\arcdeg\ in diameter divided
into 55 facets, using Gaussian source models of FWHM 0\arcsec (point
source), 36\arcsec\ and 108\arcsec, flux density cutoffs of 0.21, 0.42
and 1.7 mJy/beam respectively and $b=0.62$.  The difference between
the image in Fig.~\ref{fig:PBcor1} and that obtained with a
point-source only Clean for this dataset is shown in the left-hand
side of Fig.~\ref{fig:MSdif}.  The boxes used in the two Cleans were
identical.  Because the continuum in NGC~6503 is extended
(Fig.~\ref{fig:Ccent}), the multi-scale Clean managed that area of the
image rather better than the point-source Clean, and the difference
between the two images at this location peaks at $\sim 60 \,\mu$Jy.
The rest of the continuum image is dominated by emission from
unresolved objects (Fig.~\ref{fig:PBcor1}): for this distribution, the
benefits of a multi-scale Clean over a point-source Clean are minimal.
Nonetheless, the difference image in Fig.~\ref{fig:MSdif} shows a
circular pattern surrounding NGC~6503, with peaks in excess of $\sim
20 \,\mu$Jy.  This circular pattern is the Fourier transform of the
roughly circular hole in the data sampling due to missing spacings
less than the antenna diameter.  Multi-scale Clean appears to have
removed a systematic defect from the image, reduced the noise by a few
percent, and raised the flux density on the extended source by $\sim
9$\%.

\begin{figure}
\centering
\centerline{\resizebox{3.0in}{!}{\includegraphics{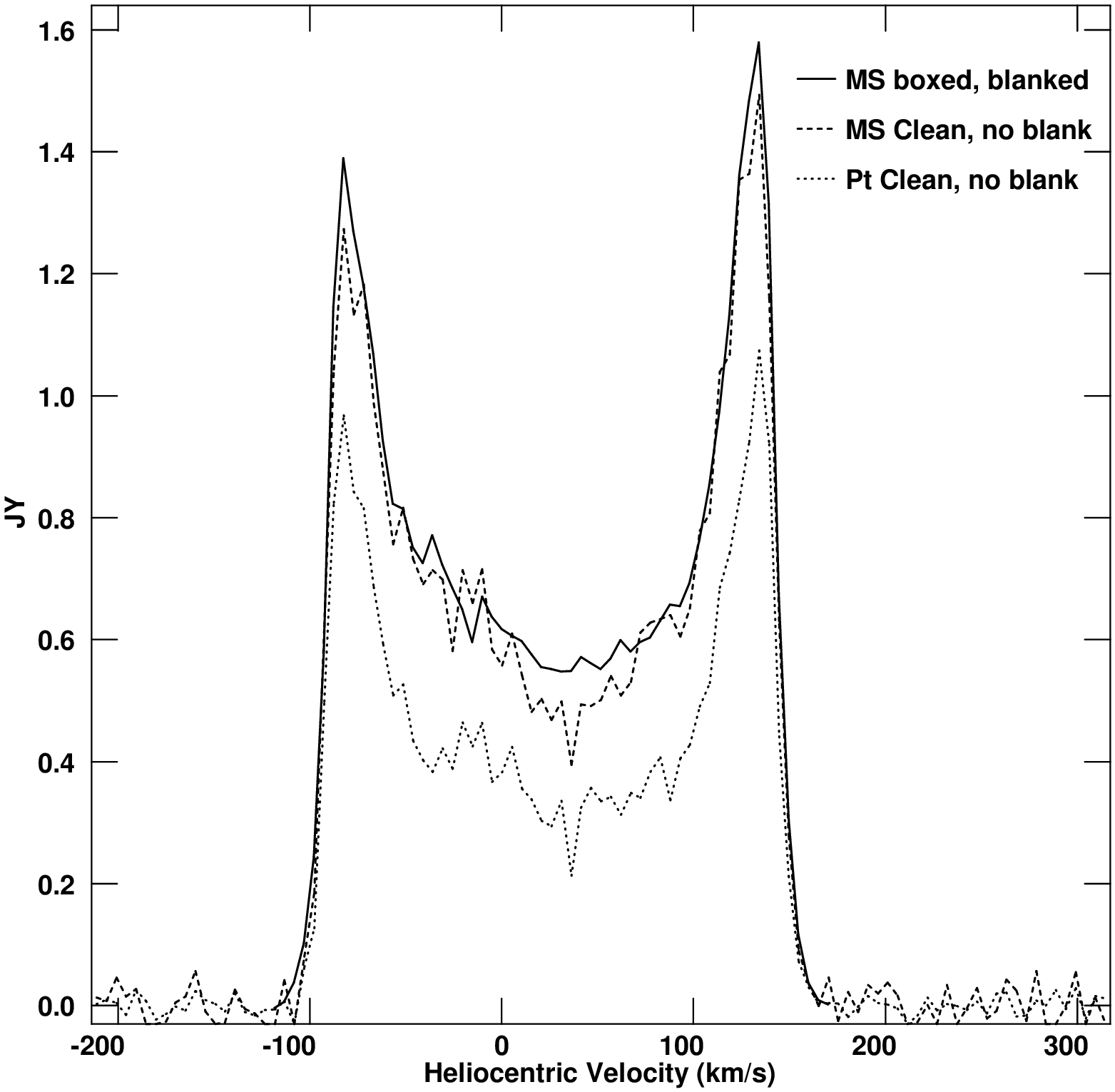}}%
 \hspace{1em}\resizebox{3.0in}{!}{\includegraphics{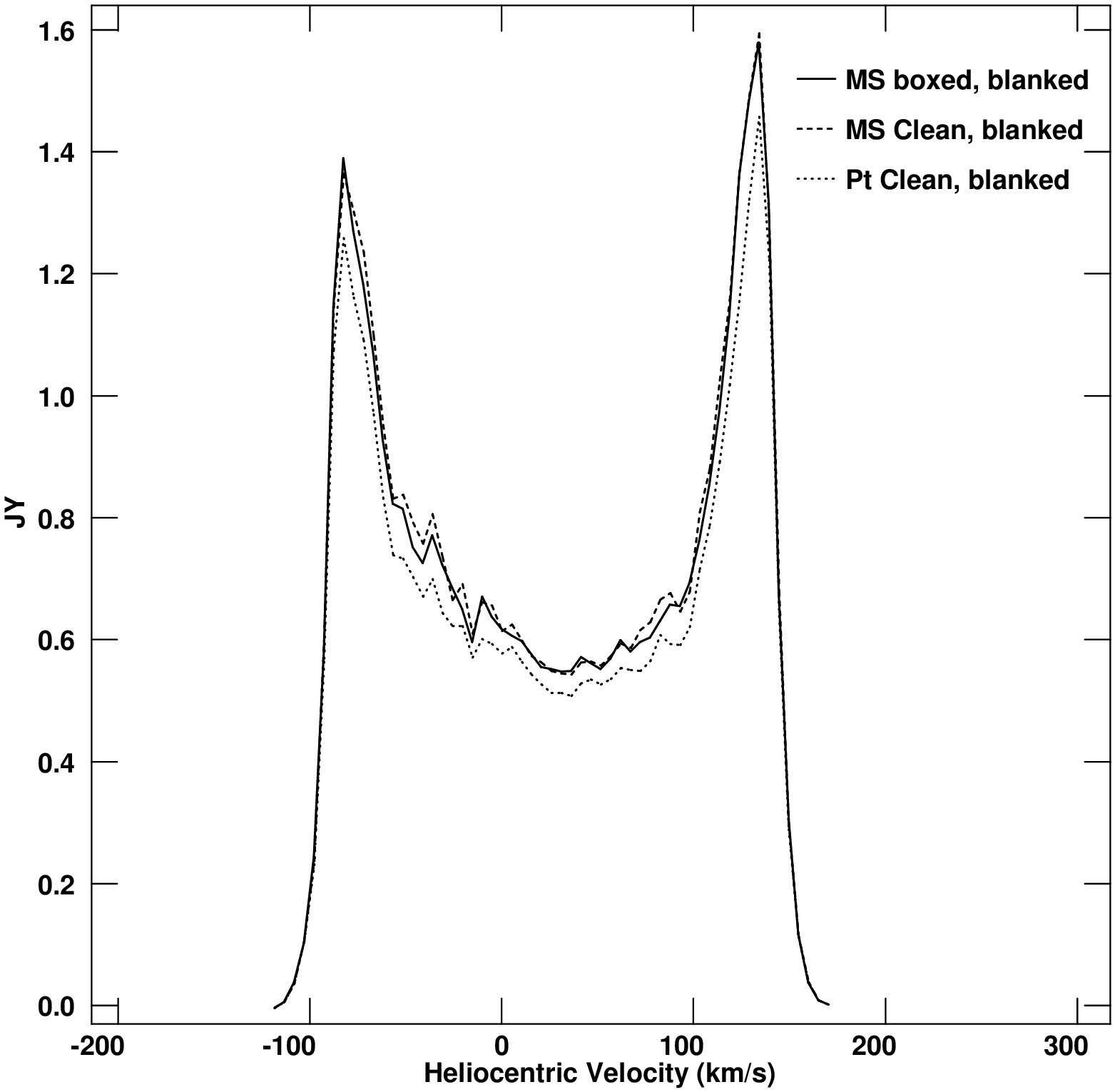}}}
\caption{Primary beam-corrected integrated spectrum of NGC~6503
  obtained using a number of Cleaning and blanking prescriptions.
  Left: the solid line is the integral over the blanked cube, made
  from a carefully boxed Clean of each channel image (reproduced from
  Fig.~\ref{fig:Ispec}).  The dashed line is the integrated spectrum
  over a rectangular region slightly larger than the extent of the
  zeroth moment image constructed with multi-scale Clean with little
  restriction of the Cleaning area.  The dotted line is the integrated
  spectrum constructed in the same way but using only point-source
  Clean components.  Right: Same as left, but only un-blanked regions
  in the carefully boxed cube are included in each profile.}
\label{fig:IspecMS}
\end{figure}

The HI in NGC~6503 was imaged in each channel in a single facet,
and deconvolved using a multi-scale Clean with 0\arcsec, 36\arcsec,
108\arcsec\ and 324\arcsec\ FHWM source models, flux density cutoffs
of 0.2, 0.55, 2.6, and 7.8 mJy/beam respectively and $b=0.62$.
Because the channels in Fig.~\ref{fig:Line4} contain extended
emission, a multi-scale Clean performs substantially better than a
point-source Clean for this cube.  As an example, multi-scale Clean
found 16\%\ more flux density in 6\%\ {\it more} iterations than the
point-source Clean at $V=98.3\,$km s$^{-1}$ using identical boxes;
see the right-hand side of Fig.~\ref{fig:MSdif}.  Relative to a
point-source Clean, the multi-scale Clean of the HI data cube has
eliminated most of the frequency-dependent ``bowl'' of negative
emission indicative of missing short spacings in the visibility data.
This is illustrated by the primary beam-corrected integrated spectra
of NGC~6503 in Fig.~\ref{fig:IspecMS}.  The dashed line in the left
panel shows the flux density of NGC~6503 Cleaned and measured in a
rectangular region encompassing the zeroth moment image
(Fig.~\ref{fig:Xmoms}).  It falls systematically below the spectrum
measured from the blanked data cube discussed in
Section~\ref{images:HI} (solid line, reproduced from
Fig.~\ref{fig:Ispec}) because negative emission from a bowl has been
included in the sum.  By contrast, the dotted line shows the same flux
density measurement as the solid one, but for a point-source Clean
data cube of the same visibility data with the same zeroth-moment,
channel-independent box area.  The dotted line falls well below the
dashed one because the bowl is much more pronounced in the
point-source Cleaned data cube than in the multi-scale Cleaned one:
33\%\ less flux density is measured from the former in this region.
In the right-hand panel of Fig.~\ref{fig:IspecMS}, the same three
integrals are plotted, but all three are restricted to the un-blanked
regions of the cube that was carefully boxed during, and carefully
blanked after, Cleaning.  The loss of flux density due to the negative
bowl has been eliminated for the multi-scale Clean (total raised from
185 to 200 Jy km s$^{-1}$), but a loss of $\sim 10$\%\ remains for the
point-source Clean (total raised from 124 to 182 Jy km s$^{-1}$).  The
cube which was Cleaned using a single large box area encompassing the
full HI galaxy, but which was carefully blanked thereafter (solid
line) has 3 Jy km s$^{-1}$ more total flux than the carefully Cleaned 
cube (dashed line), perhaps partly due to Milky Way emission and
partly to regions of emission overlooked in the careful Cleaning.  The
general conclusion of this is that careful, channel-dependent boxing
during the multi-scale imaging in not required, although some
restriction of the Clean area is always recommended.  Careful blanking
of the resulting image cube will then eliminate noise regions and some
remaining negative bowl.  Note that, if there is any negative bowl,
then the resulting total fluxes will be reduced by the fact that
signal regions are sitting inside the bowl and so have too low a
value.  Since the negative bowl problem is very much worse with the
point-source Clean, the reduction of the brightness within signal
regions remains significant.

The disparity between the left- and right-hand sides of
Fig.~\ref{fig:IspecMS} serves as a reminder that a multi-scale Clean
--- as with any other deconvolution algorithm -- does not preclude the
need for a careful analysis of the data.  Nonetheless, IMAGR's
multi-scale Clean algorithm outperforms a point-source Clean for
extended sources like NGC~6503, yielding a truer representation of the
sky brightness distribution.

\end{document}